\documentclass[10pt,twocolumn]{wlscirep} %fleqn equations move to left
\usepackage[switch]{lineno}
\usepackage[utf8]{inputenc}
\usepackage[T1]{fontenc}
\usepackage{pdflscape}   % landscape 
\usepackage{longtable}   % longtable
\usepackage{booktabs}    % \toprule, \midrule, \bottomrule, \cmidrule
\usepackage{array}       % >{\raggedright\arraybackslash}p{...}
\usepackage{makecell}    % \makecell
\usepackage{needspace}   % \needspace
\usepackage[table]{xcolor}
\usepackage{rotating}

\AtBeginDocument{%
}

\title{A general feedback mechanism unifying stochasticity, long transients, and resilience.}
\author[1,*]{Misha Chai}
\author[1]{Holger Kantz}
\affil{Max Planck Institute for the Physics  of Complex Systems, N\"othnitzer Str. 38, D 01187, Germany}
\affil[*]{Corresponding author. Email: chaimisha@pks.mpg.de}
\begin{abstract} 
Stochasticity in complex systems is commonly treated either as external noise added to deterministic dynamics or as an effective stochastic description of deterministic chaos. A framework in which stochastic dynamics arise endogenously from deterministic chaos and subsequently feed back on the deterministic dynamics, thereby fundamentally reshaping system dynamics, remains rare. Here, we develop such a framework. Through the interaction between deterministic and stochastic dynamics, alternative stable states, critical transitions, adaptive capacity, stochasticity, and resilience become dynamically interconnected within a single model. Furthermore, the framework establishes a new mechanism of long transients, driven by two key factors: (i) time-scale separation and (ii) time-delay-induced cancellation. Stochasticity plays a dual stabilizing–destabilizing role in long transients: it promotes their persistence while also facilitating their eventual escape and collapse. This dual role allows resilience to be quantified dynamically through escape and stochastic recapture dynamics. Finally, we establish the generality of the mechanism mathematically and numerically across a broad class of one-dimensional nonlinear systems, suggesting that it reflects a general dynamical principle.
\end{abstract}

\begin{document}

\flushbottom
\maketitle

\section*{Introduction}
Over the past decades, deterministic dynamics—such as attractors, bifurcations, and chaos—have been central topics in nonlinear dynamics\cite{Lorenz_1963, May_1976,Li_Yorke_1975, Feigenbaum_1978}, while stochastic theories have described fluctuations, diffusion, and noise-driven transitions\cite{Gardiner_stochastic_2009}. Considerable work has been done to bridge the gap between them. On the one hand, deterministic chaotic systems can exhibit diffusive statistics and, under appropriate conditions and rescaling, give rise to stochastic limits such as the Wiener Process or stochastic differential equations\cite{1990_Beck, ChristianBeck_1999, Holger_2001, Holger_2005, Melbourne_2011, Kelly_2017}. On the other hand, extensive studies have shown that noise can strongly modify deterministic structures, including transitions between periodic and chaotic dynamics and altering the persistence of transient states\cite{Horsthemke_1984, Lai_2003, Lai_2008, Lai_2011}. In most such descriptions, however, stochasticity is either introduced as external noise acting on an otherwise deterministic system or derived from deterministic chaos as an effective stochastic description. 

A framework in which stochastic dynamics arise endogenously from deterministic chaos, continuously interact with the deterministic dynamics, and fundamentally contribute to the resulting behavior is, however, rare. In such a framework, stochasticity no longer acts as an external noise added to a deterministic model, but as an internal component that fundamentally shapes the system's dynamics. Several fundamental challenges then arise: how does stochastic dynamics emerge endogenously from deterministic dynamics and interact with it? How does the resulting stochasticity stabilize or destabilize deterministic structures? How do stochastic and deterministic dynamics interact to generate complex behavior, and what role does stochasticity play in the emergence of long transients? These remain largely unresolved.

Meanwhile, real-world systems commonly exhibit a range of complex dynamical phenomena, including alternative stable states~\cite{2001_Scheffer}, critical transitions~\cite{Scheffer_2012}, adaptive capacity~\cite{2002_Folke_Resilience}, resilience~\cite{1973_Holling}, and stochasticity~\cite{1985_Stephen}. Although these phenomena are often studied separately, they frequently coexist and are deeply interconnected in natural systems~\cite{2009_scheffer_book}. Consequently, the other challenge is the lack of a unified framework that not only captures these phenomena within a single model, but also reveals whether and how they can arise intrinsically from a common underlying dynamical mechanism.

Among these phenomena, \textit{resilience} is of particular interest, as it extends the concept of system ``stability'' beyond the persistence of a
particular state. It is generally understood as the capacity of a system to absorb disturbances and reorganize while retaining its essential structure, function, and dynamical regime~\cite{1973_Holling,1986_Holling_Resilience,
2001_Holling_Resilience}. There are different definitions of resilience across fields\cite{1991_Pimm, 1996_Holling_resilience, 2004_Walker_resilience}. However, a general dynamical understanding of resilience remains lacking. In particular, how resilience emerges from the underlying dynamics of a system, how it interacts with deterministic and stochastic processes, and how it can be quantified dynamically remain fundamental open questions.

In this work, we develop a minimal alternating-feedback framework that addresses these questions within a single dynamical system. Rather than
introducing stochasticity as an external term, the framework allows stochastic-like dynamics to emerge endogenously from deterministic
chaos and, in turn, to feed back on the deterministic dynamics. 

First, we identify a general feedback mechanism that transforms different dynamical regimes into distinct feedback responses: chaotic dynamics in the $x$-direction generates diffusive motion in the $y$-direction with a Wiener limit, whereas periodic dynamics in the $x$-direction generates repeated slow monotonic drift in the $y$-direction, causing periodic regions to act as transient attractors. This mechanism therefore unifies stochastic-like dynamics and deterministic transient dynamics within a single dynamical framework. Importantly, stochastic and deterministic dynamics are not introduced as separate ingredients: stochastic-like dynamics emerges endogenously from deterministic chaos and subsequently feeds back on and interacts with the deterministic dynamics. As the $x$-dynamics alternates between chaotic and periodic regimes, the corresponding $y$-dynamics switches between diffusive motion and long-lived transient trapping. The interplay between these two dynamical modes drives repeated regime shifts and contributes to the emergence of long transients.

Second, we uncover a new mechanism for the emergence of long transients and identify the conditions for their emergence. Rather than merely observing long-lived transient behavior, we establish mathematically how such dynamics emerges from the internal feedback process and how transient attraction is maintained over extremely long time scales. Most importantly, we identify two key dynamical ingredients contributing to the emergence of long transients: (i) time-scale separation and (ii) time-delay-induced cancellation. Together, these provide a general explanation for how extremely long transient dynamics can arise from a deterministic system.

Third, we reveal a dual stabilizing--destabilizing role of endogenously generated stochasticity in long transients. Rather than merely observing stabilizing and destabilizing behavior, we provide a dynamical and mathematical characterization of how these opposite effects arise from the internal feedback dynamics of the system. The stochastic dynamics can promote the persistence and recovery of long-lived deterministic states, while also facilitating their eventual escape and collapse. This reveals stochasticity as an intrinsic dynamical component that regulates both persistence and loss of stability. This dual role further provides a dynamical basis for understanding resilience and, within the ecological interpretation of the framework, gives rise to a trade-off \cite{1989_Stearns_tradeoffs} between stability and biodiversity.

Fourth, we show that several phenomena that are typically studied separately---including alternative stable states, critical transitions, adaptive capacity, stochasticity, and resilience---can be dynamically connected through the same underlying feedback mechanism. Rather than introducing these properties as independent ingredients, our framework reveals how they emerge and remain dynamically interconnected within a single system. Thus, the model does not merely reproduce multiple complex-system phenomena, but provides a unified dynamical explanation for their interconnection.

Importantly, we establish that the mechanism is not specific to the logistic map used as a minimal example. The underlying alternating-feedback mechanism does not depend on a particular functional form.  The framework extends to a broader class of one-dimensional dynamical systems. We further verify this generality numerically across different one-dimensional maps; see Supplementary Fig. 1. Together, these results suggest that the mechanisms identified here reflect general dynamical principles rather than model-specific phenomena.
%---------------

%%-------------------SEC----
\section{Theoretical Framework}
\label{NLM_Theoretical_Framework}

\begin{figure*}[t]
\centering
\renewcommand{\figurename}{Fig.}
\includegraphics[width=1.0\linewidth]{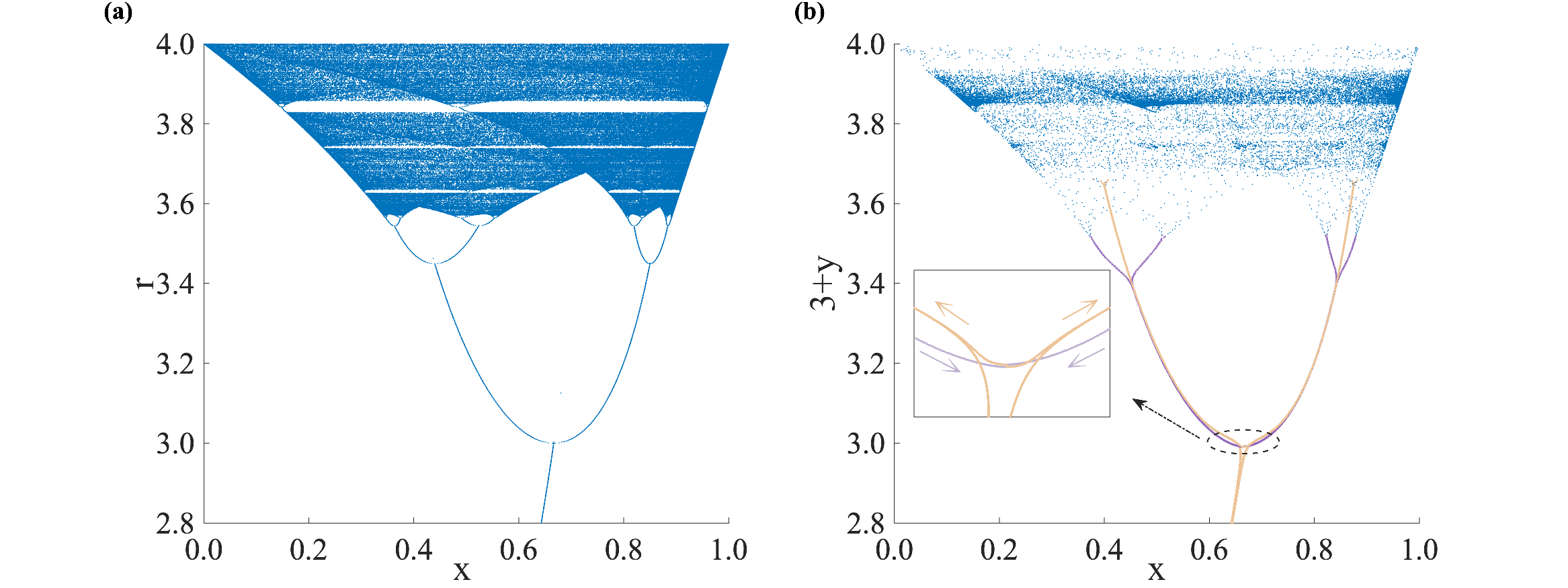}
\caption{\textbf{Phase portrait of the alternating-feedback logistic map in the $(x,y)$-plane, as defined by Eq.~\eqref{eq:map1}}. (a) $y_n = const$, achieved in the limit $a \rightarrow 0$, reducing the system to the one-dimensional logistic map; its bifurcation diagram is shown. (b) $y_n \neq const$ when $a \neq 0$ (here, $r=3 \; \text{and} \; a=0.005$). Trajectories move upward along the orange curve and downward along the purple curve. A zoomed-in view near $y=0$ is also shown.} 
\label{fig:NLM_PhaseSpace}
\end{figure*}

We begin with the logistic map\cite{May_1976}, a well-known discrete-time model that has been widely used to describe the population of a single species influenced by carrying capacity, showing a spectrum of behaviors from stable equilibrium to periodic oscillations and chaos determined by its growth rate.

The assumption of a constant growth rate in the logistic map can describe the dynamics of some simple systems or toy models properly, for example, a single-species algae population with one limiting resource \cite{1998_Scheffer_book}. In many natural systems, however, the conditions governing population
growth are themselves modified by the population dynamics. Such feedback can arise through biotic interactions, such as competition\cite{1961_Hutchinson_paradox, 2000_Chesson_diversity, 2009_Levine_diversity, 2018_Chesson_diversity, 1999_Huisman_biodiversity} and predation\cite{Holling_1959_a, 2001_Terborgh_ecological, 2011_Estes_trophic}, as well as through population-induced changes in abiotic conditions, including resource or nutrient availability\cite{2002_Vanni}, waste accumulation\cite{2018_Ratzke, 2020_Amor_transient}, and habitat quality\cite{2001_Bruno_habitat, 2007_Hastings_ecosystem}.

An ecosystem is therefore more naturally viewed as an interacting biotic--abiotic entity, in which biotic dynamics modify internal conditions and these conditions, in turn, feed back on subsequent population growth
\cite{2009_scheffer_book}. Such endogenous feedback is consistent with the view of ecosystems as complex adaptive systems, in which the system continuously modifies the conditions governing its own subsequent dynamics\cite{Levin1998}.

Accordingly, the effective growth rate $R_n$ need not remain fixed, but can evolve dynamically in response to the state of the system. We therefore write
\begin{equation}
    R_n = r + y_n,
\end{equation}
where $r$ denotes the baseline growth rate and $y_n$ represents the effect of population-induced changes in the internal conditions on population growth.

%However, in realistic systems, the growth rate is influenced by both external and internal conditions. External conditions consist of factors that are not notably affected by the system or that vary only slowly compared with the system dynamics \cite{2009_scheffer_book}. 

Since the internal conditions are modified by the population dynamics, we model the evolution of $y_n$ in terms of successive changes in $x$. The cumulative change in the population of $x$ over successive generations can be expressed as 
\begin{equation}
    \sum_{k=0}^n(x_{k+1}-x_k)=x_{n+1}-x_0.
\end{equation}
However, this expression telescopes to the net change between the initial and final states, so that the individual step-by-step changes during the evolution are no longer retained. To retain the contributions of successive population changes, we introduce a minimal alternating feedback $(-1)^j$; thus, we have
\begin{equation}
    y_{n+1}
    =
    y_0
    +a\sum_{k=0}^{n}(-1)^k(x_k-\ell),
    \label{eq:y_sum_version}
\end{equation}
where the parameter $a$ scales the influence of changes in $x$ on $y$. Here, $\ell\in I$ is an
arbitrary but fixed constant, with $I$ denoting the state space of $x$. Rather than telescoping to a single difference between the initial and final states, the alternating accumulation retains the successive pairwise differences throughout the trajectory. The fixed constant $\ell$ cancels within each pair and therefore does not affect the alternating-difference structure.

Eq.~\eqref{eq:y_sum_version} can be equivalently written in the iterative form 
\begin{equation}
      y_{n+1} =y_n+(-1)^n\cdot a(x_n-\ell),
    \label{eq:y_feedback}
\end{equation}
which describes how the feedback variable $y$ is updated at each generation. 
\iffalse
Combining two successive updates, the $y$-equation can also be written as
\begin{equation}
    y_{n+1}-y_{n-1}
    =
    \pm a(x_n-x_{n-1}),
    \label{eq:delay_form}
\end{equation}
where the sign depends on the parity of the iteration. Thus, the alternating-feedback equation also admits an equivalent time-delay interpretation in terms of successive population changes.
\fi
Thus, the alternating-feedback logistic map is
\begin{align} 
  x_{n+1} &= (r+y_n)x_n(1-x_n),  \nonumber\\ 
  y_{n+1} &= y_n+(-1)^n\cdot a(x_n-\ell).
\label{eq:map1}       
\end{align}
Here, $x_n$ denotes the population size at generation $n$, while $y_n$ represents the normalized endogenous feedback arising from population-induced changes in the internal conditions and acting on the effective growth rate. The parameter $r$ denotes the baseline growth rate in the absence of endogenous feedback, and $a$ scales the strength of the feedback between the population dynamics and the
internal conditions. For the logistic map, $I=[0,1]$, and we choose $\ell=1/2$ as the midpoint of the state space, thereby centering the feedback term. This choice is otherwise arbitrary and does not change the alternating-feedback dynamics.

Eq.~\eqref{eq:map1} defines a closed endogenous feedback loop. The population dynamics modifies the feedback variable $y_n$, while $y_n$,
through the effective growth rate $R_n=r+y_n$, in turn modifies the subsequent dynamics of $x_n$. In the limit $a\rightarrow0$, $y_n$ becomes constant and the system reduces to a one-dimensional logistic map with a fixed effective growth rate. For finite $a$, however,
$y_n$ evolves dynamically, causing the effective growth rate $R_n$ to change over time and allowing the system to move through different
dynamical regimes.

Additionally, the phase space of our map is the product of the $(x,y)$-plane and the set $\{-1, 1\}$. The diagram in the $(x,y)$-plane shown in this paper represents a projection where the two layers, $\{-1\}$ and $\{1\}$, are both mapped onto the same $(x,y)$-plane. As a result, the dynamics may appear non-unique, or different trajectories may appear to intersect in this graphical projection. However, this apparent non-uniqueness is purely a consequence of the graphical projection. In the full phase space, where the additional discrete component $\{-1, 1\}$ is retained, the trajectories remain distinct and therefore unique. 
%Further details are provided in later sections.

%The $x$-dynamics remain consistent with Eq.~(\ref{eq:map1}). Therefore, given $x_1$ and $y_1$ as initial conditions, the whole trajectory is uniquely determined. This representation of our map reveals that its phase space is the direct product of the $(x,y)$-plane and the set $\{-1,1\}$. In the graphical representations of the attractor, we will focus solely on projections onto the $(x,y)$-plane. However, the fact that there is the additional coordinate $sign=\pm1$ explains why the dynamics appears to be not unique in such a projection. 

%%--------------------SEC----
\section{System Dynamics Analysis}
\label{NLM_System_Dynamics_Analysis}
When $y_n=\mathrm{const.}$ (achieved in the limit $a\to0$), Eq.~\eqref{eq:map1} reduces to a one-dimensional logistic map. The bifurcation diagram then reveals the dynamics from stable equilibrium through period-doubling bifurcations and eventually into chaotic dynamics as $r$ increases in the 1D logistic map, as shown in \autoref{fig:NLM_PhaseSpace}(a).

When $y_n\neq\mathrm{const.}$ (i.e., $a\neq0$), the essential difference is that the effective growth rate $R_n=r+y_n$ now evolves dynamically with $y_n$. Consequently, the trajectory moves
through different regions of the underlying logistic-map bifurcation diagram. As shown in \autoref{fig:NLM_PhaseSpace}(b), $y$ drifts upward along the orange curve and downward along the purple curve through the period-doubling bifurcation region. Meanwhile, it also drifts through the chaotic and periodic window regions. Correspondingly, $x$ alternates between chaotic and periodic regimes. The system, therefore, exhibits intermittency, with trajectories displaying varied intermittent behaviors as $y$ evolves. Moreover, since $R_n=r+y_n$ is not restricted to being less than or equal to 4, and if $R_n>4$, then the $x$-dynamics map trajectories outside the interval $[0,1]$. Consequently, the trajectories escape towards infinity ($\infty$), implying that the system also exhibits transient behavior. We therefore terminate the iteration, considering the trajectory as having escaped. Overall, the system exhibits a rich interplay of intermittency and transient dynamics, driven by the drift and the unbounded range of $y$. 

The behavior of the variable $y_n$ can be very different, depending on whether the $x_n$ that are added up are chaotic or periodic.

\subsection{Chaotic $x$-Dynamics: Diffusive Motion in $y$}
\label{subsec:chaotic_x}

When $x_j$ evolves in a chaotic regime satisfying sufficiently strong mixing conditions, the corresponding $y$-dynamics exhibits diffusive behavior. Under diffusive rescaling, the normalized process converges weakly to a Wiener process. A detailed derivation is provided in \hyperref[box:chaotic_diffusive]{Appendix.~\ref*{box:chaotic_diffusive}}. Consequently, for large $n$,
\begin{equation}
\left\langle \left(y_n-\langle y_n\rangle\right)^2 \right\rangle \sim Dn, \qquad n\to\infty. 
\label{eq:NLM_BM}    
\end{equation}
The corresponding effective diffusion coefficient is
\begin{equation}
D
=
a^2
\left[
C(0)
+
2\sum_{k=1}^{\infty}
(-1)^k C(k)
\right],
\end{equation}
where $C(k)$ denotes the correlation function of the centered chaotic process, with $C(0)=\sigma^2$.

This result is not specific to a particular chaotic map, but applies to a broad class of strongly mixing chaotic dynamics satisfying the above conditions.

Moreover, owing to the alternating sign $(-1)^j$, 
the bounded term $y_n^{(\mathrm{bnd})}$ remains uniformly bounded for all $n$, as proved in \hyperref[box:chaotic_diffusive]{Appendix.~\ref*{box:chaotic_diffusive}}, Eq.~\eqref{eq:bounded_term}. The constant $\ell$ affects only this bounded term, which is negligible on diffusive scales and therefore does not alter the diffusive behavior. Hence, $\ell$ can be any fixed constant within the state space $I$ without changing the Wiener limit.

In summary, two key ingredients contribute to the emergence of diffusive motion in $y$-direction: (i) the alternating factor $(-1)^n$ keeps the bounded term $y_n^{(\mathrm{bnd})}$ uniformly bounded, so that it becomes negligible under diffusive rescaling; and (ii) sufficiently strong chaotic mixing, together with the corresponding functional central limit theorem, causes the diffusive term $y_n^{(\mathrm{diff})}$ to exhibit diffusive scaling and converge weakly
to a Wiener process. 

%For the strongly mixing regimes considered here, the nonzero-lag correlations decay rapidly and their overall contribution is small compared with the zero-lag variance. Hence, to leading order,
%\begin{equation}
%D\approx a^2\sigma^2.
%\end{equation}

%%---
\subsection{Periodic $x$-Dynamics: Transient Attractor}
\label{subsec:periodic_x}
When $x_n$ evolves in a periodic regime, the corresponding $y$-dynamics differs qualitatively from that in the chaotic regime. Instead of diffusive motion, $y_n$ exhibits monotonic dynamics in the periodic region, causing periodic regions to act as transient attractors. We consider two representative cases: periodic windows and the period-doubling bifurcation region.

\begin{figure*}[t]
\centering
\renewcommand{\figurename}{Fig.}
\includegraphics[width=1.0\linewidth]{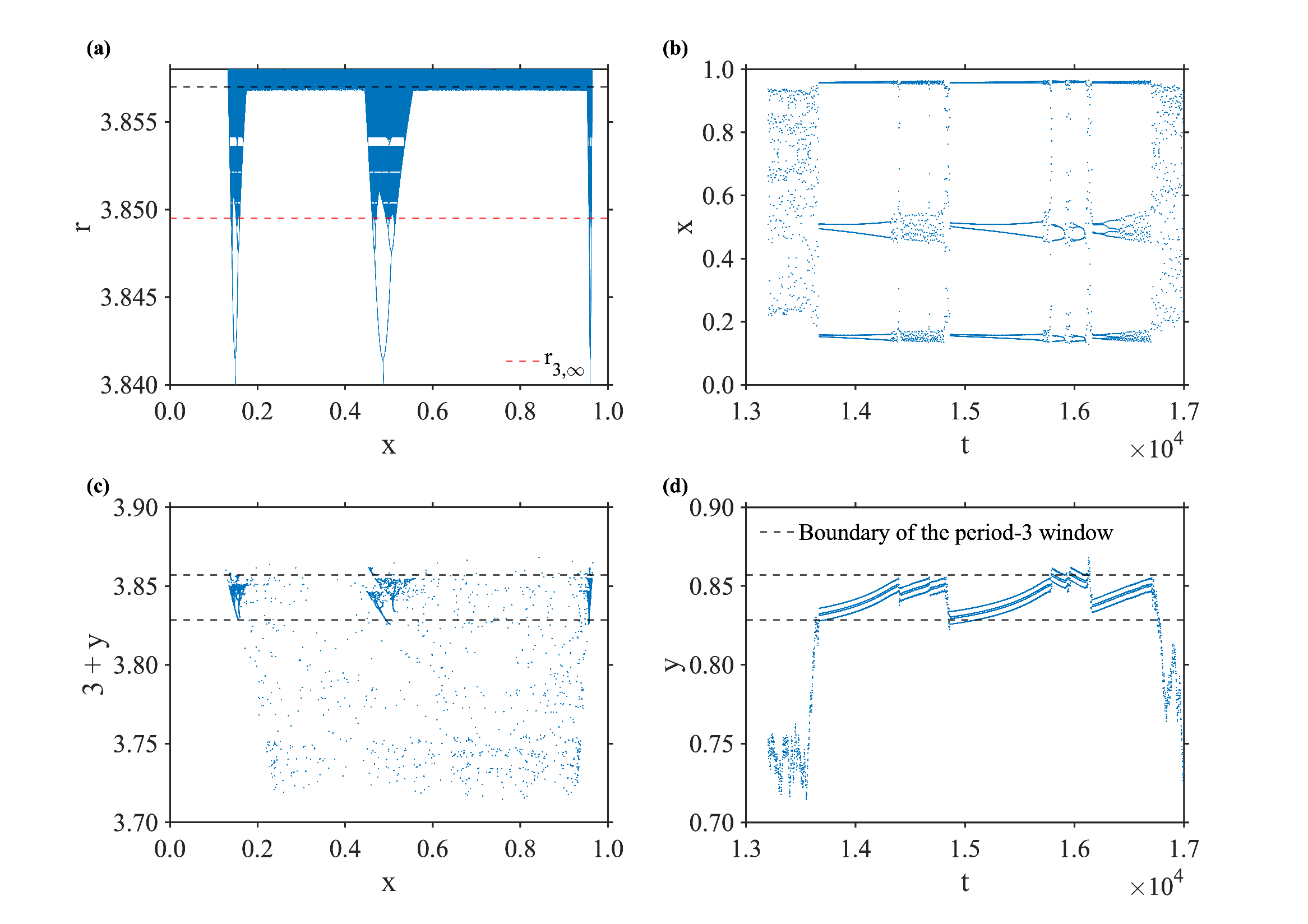}
\caption{\textbf{Dynamics of $(x,y)$ near the period-3 window.} Initial conditions are chosen randomly from $(x_0,y_0) \in (0, 1) \times (0, 1)$, with $r=3$ and $a=0.01$. Black dashed lines indicate the approximate lower ($R\approx3.8284$) and upper ($R\approx3.857$) boundaries of the period-3 window. (a) Bifurcation diagram near the period-3 window. The red dashed line indicates the period-$3 \cdot 2^\infty$ orbit at $r_{3, \infty} \approx 3.8495$. (c) Trajectories in the $(x,y)$-plane concentrated within the period-3 window. (b, d) Corresponding time series of $x$ and $y$, respectively.} 
\label{fig:NLM_p3_details}
\end{figure*}

\subsubsection{Periodic Windows}
When $x_n$ evolves within a period-$L$ window (e.g., the classic period-3 window occurring over the interval $R_n \in (1+\sqrt{8}, 3.857)$\cite{Strogatz_2015}), the $y$-dynamics exhibits quasi-periodic behavior owing to the slow drift in $y$. More precisely, while $x_n$ evolves along a quasi-period-$L$ orbit, $y_n$ evolves along a corresponding quasi-period-$2L$ orbit. To describe the slow drift between successive cycles, we introduce a double index notation: let $(X_{i,j}, Y_{i, j})$ denote the $j$-th point ($j=0, 1, ..., 2L-1$) of the $i$-th quasi-cycle. $(X_{i,j}, Y_{i, j})$ corresponds to a quasi-period-$2L$ orbit. Thus, the difference between successive quasi-cycles can be quantified as: 
\begin{align}
    X_{i+1,j} &= X_{i,j} + a\delta_i f(X_{i+1,j-1}) && (i,j)\neq(0,0),
    \nonumber \\
    Y_{i+1,j} &= Y_{i,j} + a\delta_i && (\forall\; i, j),
    \label{eq:NLM_periodicity}
\end{align}
where 
\begin{equation}
    \delta_i=\sum_{j=0}^{2L-1}(-1)^jX_{i, j},
\end{equation}
and
\begin{equation}
 f(X_{i+1,j-1})=X_{i+1,j-1}(1-X_{i+1,j-1}).  
\end{equation}
A detailed mathematical proof is provided in \hyperref[box:periodic_ballistic]{Appendix.~\ref*{box:periodic_ballistic}}. 

More importantly, the coupling between $y_n$ and $x_n$ induces a monotonic drift in the $Y$-direction. The direction of this drift is determined by the initial sign of $\delta_1$, where $\delta_1$ denotes the value of $\delta_i$ when $Y$ first reaches the boundary of the periodic window:
\begin{equation}
\operatorname{sgn}(\delta_i) = \operatorname{sgn}(\delta_1) \ \text{for all } i \geq 1.  
 \label{eq:NLM_quasi_condition2}
\end{equation}
Once the sign of $\delta_1$ is fixed, the direction of the drift in the $Y$-direction is uniquely determined. This produces a monotonic drift in the $Y$-direction within the periodic region, pushing $Y_{i,j}$ toward the boundary of the periodic region. 

The strength of this drift, however, differs depending on whether $L$ is odd or even. For odd $L$, the alternating factor $(-1)^j$ produces pairwise cancellation between corresponding points of the two successive quasi-period-$L$ segments, making $\delta_i$ very small. Consequently, for sufficiently small $a$, the variation in the $Y$-direction from one quasi-cycle $i$ to the next $i+1$ satisfies
\begin{equation}
     |Y_{i+1,j}-Y_{i,j}|\ll 1.
    \label{eq:NLM_quasi_condition1}
\end{equation}
As $Y$ slowly and monotonically sweeps through the periodic region, it successively gives rise to a large number of transient quasi-period-$2L$ orbits $(X_{i,j}, Y_{i,j})$ within the period-$L$ window. 

For even $L$, by contrast, since $(-1)^j=(-1)^{L+j}$, the pairwise cancellation between two successive quasi-period-$L$ segments in the $X$-direction does not occur. so that $|\delta_i|$ is generally not small. Consequently, the cycle-to-cycle displacement $|a\delta_i|$ in the $Y$-direction is relatively large, causing trajectories to pass through even-period windows more rapidly. Even-period windows can therefore still act as transient attractors, but their transient attraction is considerably weaker than that of odd-period windows.
%Nevertheless, since the sign of \(\delta_i\) remains preserved, the \(Y\)-dynamics still exhibits a monotonic drift.

Once $Y_{i,j}$ moves slightly outside the boundary of the periodic window, $|a\delta_i|$ quantifies the escape step in the $Y$-direction---the per-cycle displacement that drives the trajectory across the boundary of the basin of attraction. Subsequent re-entry, when it occurs, is instead enabled by diffusive motion in the neighboring chaotic region. Specifically, if $a$ is too large, the escape step $|a\delta_i|$ is correspondingly large, pushing trajectories far outside the basin of attraction. This makes it difficult to re-enter and be re-captured by the attractor. Conversely, if $a$ is small, the escape step remains tiny, so trajectories stay close to the basin boundary and are more readily pulled back into the attractive region. In this sense, \textbf{the parameter $a$ scales the effective resilience of the transient attractor}---quantifying how readily trajectories are retained within or expelled from the basin of attraction. The smaller $a$ is, the greater the basin of resilience: trajectories are more readily retained within the basin of attraction and are less likely to be expelled from it.

%When $Y$ reaches the boundary of a periodic region and moves slightly outside it, the diffusive motion rapidly drives $Y$ back into the periodic region, resulting in cyclic dynamics. These dynamics are characterized by Eq.~\eqref{eq:NLM_quasi_condition1} and Eq.~\eqref{eq:NLM_quasi_condition2}.

The underlying mechanism remains qualitatively unchanged for different values of $r$ and signs of $a$. For illustration, here we take $a>0$ and $r=3$ as an example. The dynamics can be classified into two cases: (i) $Y_{i,j}$ hits the upper boundary of a periodic window, and (ii) $Y_{i,j}$ hits the lower boundary.

\textbf{Case I: Upper boundary.}
When $Y_{i,j}$ hits the upper boundary of a periodic window, its subsequent evolution depends on the sign of $\delta_1$:

If $\delta_1>0$, $Y_{i,j}$ drifts upward into the adjacent chaotic region and begins a diffusive motion in the $Y$-direction. This diffusion allows $Y_{i,j}$ either to escape the phase space through the boundary $R=4$ or, alternatively, to re-enter the periodic window with $\delta_1<0$.

If $\delta_1<0$, $Y_{i,j}$ drifts downward into the periodic window, where successive updates with $Y_{i+1,j}<Y_{i,j}$ push it monotonically towards the lower boundary. Once $Y_{i,j}$ moves slightly outside the lower boundary of the periodic window, the near cancellation in $\delta_i=\sum_{j=0}^{2L-1}(-1)^jX_{i,j}$ breaks down, leading to the rapid growth of $|\delta_i|$. This, in turn, irreversibly drives $Y_{i+1,j}=Y_{i,j}+a\delta_i$ to rapidly escape from the periodic window and enter the chaotic region. Subsequently, the diffusive motion in the $Y$-direction within the chaotic region again provides an opportunity for $Y_{i+1,j}$ to re-hit the periodic window, producing repeated cycles of transient trapping in the period-$L$ window. 

Consequently, \textbf{the periodic window act as a transient attractor} in this process.

\textbf{Case II: Lower boundary.}
Similarly, when $Y_{i,j}$ hits the lower boundary of a periodic window:

If $\delta_1>0$, $Y_{i,j}$ drifts upward into the periodic window, where successive updates with $Y_{i+1,j}>Y_{i,j}$ push it monotonically towards the upper boundary. Once $Y_{i,j}$ moves slightly outside the upper boundary, the near cancellation in $\delta_i$ breaks down, leading to the rapid growth of $|\delta_i|$. This, in turn, drives $Y_{i+1,j}=Y_{i,j}+a\delta_i$ quickly into the chaotic region. Subsequently, diffusive motion in the $Y$-direction within the chaotic region enables repeated re-entries into the periodic window, leading to recurrent transient trapping. The periodic window thus also acts as a transient attractor. 

If $\delta_1<0$, $Y_{i,j}$ drifts downward into  the adjacent chaotic region and undergoes diffusive motion in the $Y$-direction, which further gives $Y_{i,j}$ the opportunity to re-hit the periodic window with $\delta_1>0$.

The key difference between the upper and lower boundaries lies in their relative attractiveness: \textbf{the lower boundary is more attractive than the upper}. This is because trajectories approaching the lower boundary converge to a quasi-period-$2L$ orbit, while trajectories approaching the upper boundary converge to a long quasi-period-$2L \cdot 2^n$ orbit ($n\rightarrow \infty$). For a longer quasi-periodic orbit, the corresponding $X$-points involved in the pairwise cancellation are separated by more iterations, allowing more drift in $Y$ to accumulate between them. Consequently, the accumulated drift in $Y$ becomes larger over the longer orbit, weakening the pairwise cancellation and thereby increasing
$|\delta_i|$. This, in turn, produces a larger drift in the $Y$-direction near the upper boundary. Consequently, for small $a$, since $Y_{i+1, j}-Y_{i,j}=a\delta_i$, the lower boundary is more attractive than the upper boundary. 

In summary, three key ingredients contribute to the long-lived transient attraction associated with the periodic window: (i) the alternating factor $(-1)^n$ produces strong pairwise cancellation between corresponding points of the quasi-period-$2L$ orbit when $L$ is odd, making $|\delta_i|$ very small. For sufficiently small $a$, this leads to a very small cycle-to-cycle displacement $|a\delta_i|$ in the $Y$-direction and hence a very slow drift through the periodic region; (ii) the sign of $\delta_i$ remains unchanged over successive quasi-cycles, so that the slow drift in the $Y$-direction is monotonic; and (iii) when the trajectory leaves the periodic window and enters a neighboring chaotic region, the resulting diffusive motion in the $Y$-direction provides repeated opportunities for the trajectory to re-enter and be recaptured by the periodic region. Together, these mechanisms generate recurrent long-lived transient trapping, causing the periodic region to act as a long-lived transient attractor.

\textbf{Example.}
Here we take the classic period-3 window as an example. The bifurcation diagram near the period-3 window is shown in \autoref{fig:NLM_p3_details}(a). The period-3 window lies approximately in the interval $R_n \in (1+\sqrt{8}, 3.857)$ (between the black dashed lines), where $R = 1+\sqrt{8} \approx 3.8284$ marks the saddle-node bifurcation that creates a period-3 orbit, and $R \approx 3.857$ is where the 3-band-chaotic attractor undergoes a boundary crisis, merging into broader chaos. However, the dynamics within this window are far richer than just a period-3 regime: the period-3 orbit undergoes a period-doubling bifurcation ($3 \rightarrow 6 \rightarrow 12 \rightarrow ... \rightarrow 3 \cdot 2^n$), accumulating at $R_{3, \infty} \approx 3.8495$ (red dashed line) where the orbit reaches period $3 \cdot 2^\infty$. Beyond this point, the attractor becomes a 3-band-chaotic attractor that still contains infinitely many periodic windows---each a miniature copy of the full logistic map bifurcation diagram---until $R \approx 3.857$, where the system enters chaos.

Starting from a random initial point $(x_0,y_0) \in (0, 1) \times (0, 1)$, with $r=3$ and $a=0.01$, the dynamics of $(x,y)$ are shown in \autoref{fig:NLM_p3_details}. In panel (c), the trajectory in the $(x,y)$-plane shows quasi-periodic behavior concentrated within the period-3 window (between the black dashed lines), forming three distinct branches that correspond to the period-3 orbit and its nearby bifurcation structure. Since the phase space is the product of the $(x,y)$-plane and the set $\{-1, 1\}$, the diagram in the $(x,y)$-plane represents a projection where the two layers, $\{-1\}$ and $\{1\}$, are both mapped onto the same $(x,y)$-plane. As a result, the dynamics appear non-unique in this graphical projection: although the figure shows three distinct branches, each of these branches is, in fact, composed of two overlapping sub-branches associated with $\{-1\}$ and $\{1\}$, respectively. Panels (b) and (d) show the time series of $x$ and $y$. In panel (d), when $Y_{i, j}$ hits the lower boundary of the period-3 window with $\delta_i>0$, it drifts upward until reaching the upper boundary. As $Y_{i, j}$ moves slightly outside the upper boundary, the rapid growth of $|\delta_i|$ irreversibly drives it to escape rapidly from the period-3 window into the chaotic region. Subsequently, the diffusive motion in the $Y$-direction allows $Y_{i,j}$ to re-hit the lower boundary of the period-3 window with $\delta_i>0$, repeating the cycle. 

Similarly, when $Y_{i, j}$ hits the upper boundary of the period-3 window with $\delta_i<0$, for sufficiently small $a$, it drifts downward through the complex bifurcation diagram within the period-3 window. During this process, the trajectory may pass through increasingly long periodic orbits, such as those with period $3\cdot 2^n$ for large $n$, or enter a
band-chaotic region. In either case, the near cancellation in $\delta_i$ becomes weaker, causing $|\delta_i|$ to increase and eventually driving the trajectory out of the period-3 window into the chaotic region. The period-3 window acts as a transient attractor in these processes. Meanwhile, in panel (b), $x$ exhibits chaotic and periodic behaviors, respectively, i.e., intermittency.

As $a \rightarrow 0$, once $Y_{i, j}$ enters the periodic window, we have $Y_{i+1,j}-Y_{i, j}=a\delta_i \rightarrow 0$. This indicates that the drift through the periodic window becomes arbitrarily slow, and the corresponding lifetime that trajectories spend in the periodic window diverges. Consequently, the boundary of the periodic window effectively acts as a stable attractor in this limit. However, for any finite $a \neq 0$, the lifetime within the periodic window---though extremely large---remains finite. Therefore, the periodic window remains a transient attractor.

\begin{figure*}[t] %[h!]
\centering
\renewcommand{\figurename}{Fig.}
\includegraphics[width=1.0\linewidth]{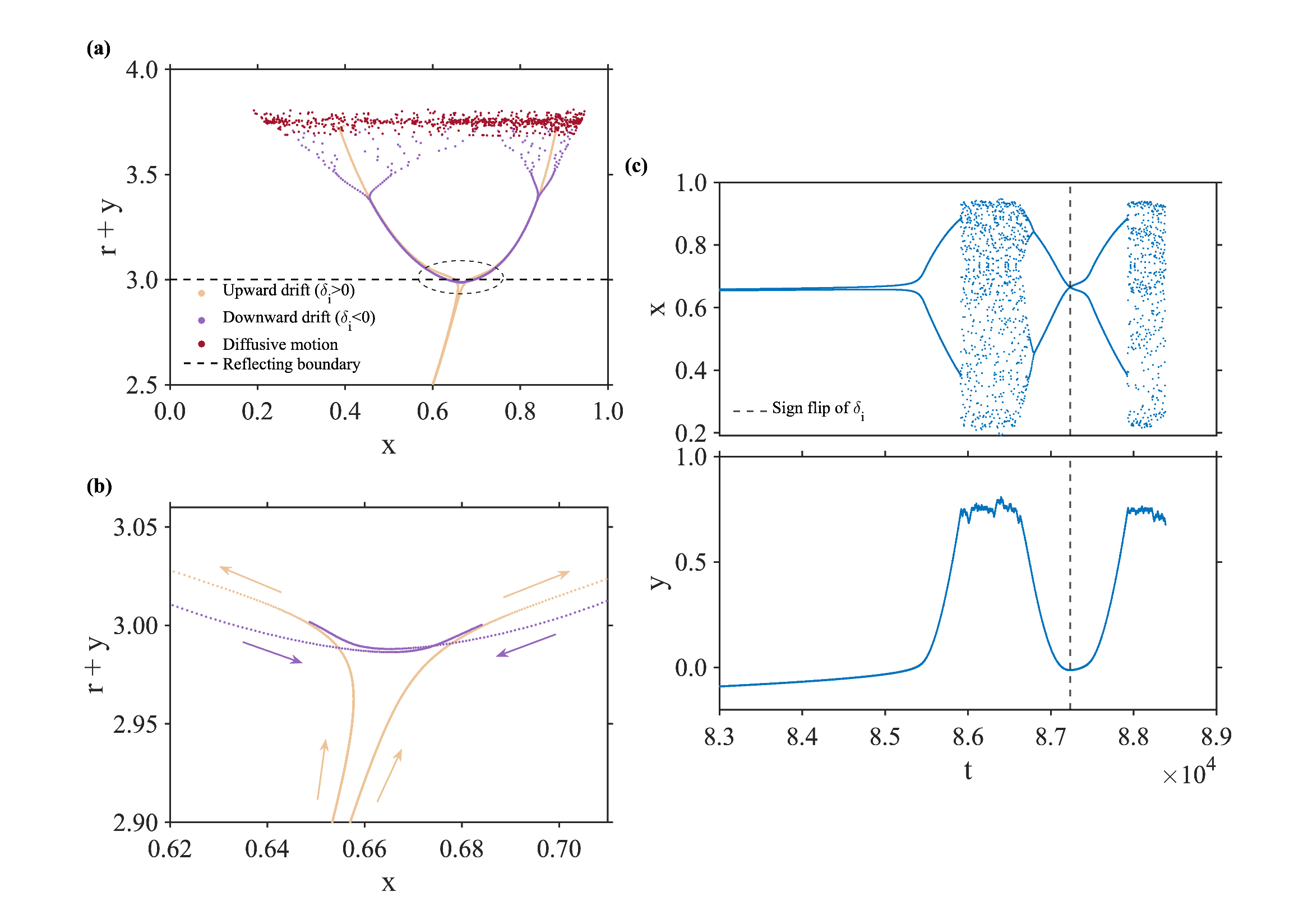}
\caption{\textbf{Dynamics of $(x,y)$ near the period-doubling bifurcation region.} Initial conditions are chosen randomly with $x_0 \in (0, 1)$ and $y_0 \in (-1, 0)$, with $r=3$ and $a=0.01$. (a) Trajectories in the $(x,y)$-plane near the period-doubling bifurcation region. Black dashed line at $R=3$ is the reflecting boundary. Trajectories move upward along the orange curve and downward along the purple curve. (b) Zoomed-in view near the reflecting boundary. (c) Corresponding time series of $x$ and $y$. The black dashed line marks the time at which the trajectory passes the reflecting boundary, resulting in a sign flip of $\delta_i$.} 
\label{fig:NLM_PDB_details}
\end{figure*}
%%---
\subsubsection{Period-Doubling Bifurcation Region} %Monotonic Drift
When $x_j$ evolves within the period-doubling bifurcation region---specifically, for $R_n\in (3, r_\infty)$ where $r_\infty \approx 3.569946$\cite{Strogatz_2015}---the resulting $y$-dynamics are quasi‑periodic owing to the slow drift in $y$. More precisely, while $X_{i,j}$ evolves within a quasi-period-$2^n$ orbit, $(X_{i,j}, Y_{i,j})$ evolves along a corresponding quasi-period-$2\cdot2^n$ orbit. 

Owing to this quasi-periodicity, for sufficiently small $a$, the same cycle-to-cycle framework developed in \hyperref[box:periodic_ballistic]{Appendix.~\ref*{box:periodic_ballistic}} can be applied here as well. In particular, the $Y$-dynamics exhibits a monotonic drift, whose direction is determined by the initial sign of $\delta_1$.

Compared with odd-period windows, however, the period-doubling bifurcation region exhibits weaker transient attraction. In odd-period windows, strong pairwise cancellation between the two successive quasi-period-$L$ segments makes $|\delta_i|$ very small, resulting in a very slow drift in the $Y$-direction. Here, $L=2^n$ is even, so this cancellation does not occur and the cycle-to-cycle displacement $|a\delta_i|$ is generally larger. The trajectory therefore moves through the region more rapidly, although the monotonic drift still allows the period-doubling bifurcation region to act as a transient attractor.

We still take $a>0$ as an example. When $Y_{i,j}$ first hits the upper boundary of the period-doubling bifurcation region, i.e., the period-doubling accumulation point near $R=r_\infty$, its subsequent evolution depends on the sign of $\delta_1$.

If $\delta_1>0$, $Y_{i, j}$ drifts upward into the adjacent chaotic region and begins diffusive motion in the $Y$-direction. This diffusion provides $Y_{i, j}$ the possibility to escape the phase space through the boundary $R=4$, or, alternatively, gives it the opportunity to re-hit the period-doubling region with $\delta_1<0$.

If $\delta_1<0$, $Y_{i,j}$ drifts downward through a sequence of reverse period-doubling bifurcations. As the higher-period branches progressively approach and merge into lower-period orbits, the trajectory can continuously follow the changing periodic structure. Since the sign of $\delta_i$ is preserved across successive quasi-cycles, the $Y$-dynamics remains monotonic, allowing the
trajectory to continue through the reverse period-doubling sequence toward the boundary $R=3$.

Near the boundary $R=3$, the quasi-period-2 orbit converges as $X_{i,j+1}-X_{i,j} \rightarrow 0$. Due to the finite drift in $y$, however, $X_{i,j+1}\neq X_{i,j}$ for $a \neq 0$.  Nevertheless, their separation becomes sufficiently small near $R=3$ that the $X_{i,j}$ can be approximated as a fixed point; thus, $(X_{i,j}, Y_{i,j})$ corresponds to a quasi-period-$2\cdot 2^0$ orbit. Consequently, $\delta_i=\sum_{j=0}^{1}(-1)^jX_{i,j}=X_{i,0}-X_{i,1}$. Before the two $X$-branches intersect, $\delta_i<0$, i.e., $X_{i,0}<X_{i,1}$. Once the two branches cross, their ordering is
reversed, giving $\delta_{i+1}>0$, i.e., $X_{i+1,0}>X_{i+1,1}$. Consequently, $Y_{i+2,j}>Y_{i+1,j}$, and the direction of the $Y$-drift reverses from downward to upward.

The subsequent upward dynamics is qualitatively different from the preceding downward motion. As $Y_{i,j}$ drifts upward away from $R=3$, the difference between successive $X$-values, $|\delta_i|=| \sum_{j=0}^1 X_{i,2j+1}-X_{i,2j}|$, increases, producing a larger cycle-to-cycle displacement $|a\delta_i|$ in the $Y$-direction. The upward drift therefore accelerates. Although the corresponding period-2 orbit loses stability at the first period-doubling bifurcation, the trajectory does not leave the quasi-period-2 structure immediately.
Because $y_n$ changes dynamically, the instability takes finite time to develop, while the accelerating $Y$-drift carries the trajectory rapidly through the period-doubling region. Consequently, the trajectory enters the chaotic region without successively following the period-doubling bifurcation.

In this sense, \textbf{$R=3$ acts as a reflecting boundary}, where the sign flip of $\delta_i$ reverses the direction of the $Y$-drift. \textbf{The period-doubling bifurcation region acts as a transient attractor}: trajectories entering the region can remain trapped for
finite times while drifting downward through the reverse
period-doubling sequence, reversing direction near $R=3$, and then rapidly moving upward with a quasi-period-$2$ structure into the adjacent chaotic region. Diffusive motion in the neighboring chaotic region can subsequently return trajectories to the period-doubling region, producing repeated transient trapping.

Additionally, for $R_n \in (2,3)$, if the initial condition $\delta_1<0$, $Y_{i,j}$ drifts downward to minus infinity. If the initial condition $\delta_1>0$, $Y_{i, j}$ drifts upward until it has the chance to escape through the boundary $R=4$. 

\textbf{Example.} Starting from a random initial point $x_0 \in (0, 1)$ and $R_0 \in (2,3)$ (in this case, $r=3$ and $a=0.01$), the dynamics of $(x,y)$ are shown in \autoref{fig:NLM_PDB_details}. Panels (a) and (b) show the $(x,y)$-plane. In panel (a), as $\delta_i>0$, $Y_{i, j}$ drifts upward along the orange curve until reaching the chaotic region (red dots). Within this region, diffusive motion gives $Y_{i,j}$ the opportunity to re-hit the period-doubling region with $\delta_i<0$. Subsequently, $Y_{i,j}$ drifts downward along the purple curve, undergoing a reverse period-doubling bifurcation until it reaches the reflecting boundary $R=3$, where
$\delta_i$ changes sign. The drift then reverses, and $Y_{i,j}$ moves upward again along the orange curve. Panel (b) shows a zoomed-in view near the reflecting boundary, where $X_{i, j}$ can be approximated as a fixed point. Thus, $(X_{i, j}, Y_{i, j})$ corresponds to a quasi-period-2 orbit with $\delta_i=X_{i,0}-X_{i,1}$. Initially, $\delta_i<0$, i.e., $X_{i,0}<X_{i, 1}$,  until $X_{i,0}$ and $X_{i,1}$ intersect. This intersection in the $X$-direction leads to $X_{i+1,0}>X_{i+1, 1}$, resulting in $\delta_{i+1}>0$. Consequently, $Y_{i+2, j}>Y_{i+1, j}$, causing $Y_{i, j}$ to drift upward again along the orange curve and repeat similar dynamics. The time series of $x$ and $y$ are shown in panel (c). As seen here, $Y_{i, j}$ drifts upward until entering the chaotic region. Within this region, diffusive motion provides $Y_{i,j}$ the possibility to re-enter the period-doubling bifurcation region with $\delta_i<0$. Subsequently, $Y_{i,j}$ drifts downward until it reaches the reflecting boundary $R=3$. At this boundary, a sign flip occurs ($\delta_i>0$, black dashed line in panel (c)), causing $Y_{i,j}$ to drift upward again, re-entering the chaotic region and repeating similar dynamics. Meanwhile, $x$ alternates between chaotic and periodic behaviors. The apparent intersections between the orange and purple curves in panels (a) and (b) arise solely from the graphical projection of the two layers, ${-1}$ and ${1}$, onto the same $(x,y)$-plane. In the full state space, the trajectories remain distinct and therefore unique.

Both the periodic window and the period-doubling bifurcation region act as transient attractors. However, the essential differences are as follows: (i) the periodic window attracts trajectories through both its upper and lower boundaries, whereas the period-doubling bifurcation region can be entered only through its
upper boundary near $r_\infty$; and (ii) the lower boundary is more attractive than the upper boundary because the corresponding $|\delta_i|$ is smaller; (iii) odd-period regions exhibit stronger transient attraction than even-period regions because the pairwise cancellation in \(\delta_i\) leads to a slower drift in the \(Y\)-direction. Based on these three features, \textbf{the periodic window with odd $L$ exhibits stronger transient attraction than the period-doubling bifurcation region}.

% In the whole paper, initial conditions are chosen randomly.
%Hence, for small $a$ (slow change of growth rate), the resulting phase space diagram strongly resembles the bifurcation diagram of the logistic map (\autoref{fig:fig2}b).

%%-------------------SEC---------
\begin{figure*}[t]%[h!]
\centering
\renewcommand{\figurename}{Fig.}
\includegraphics[width=0.9\linewidth]{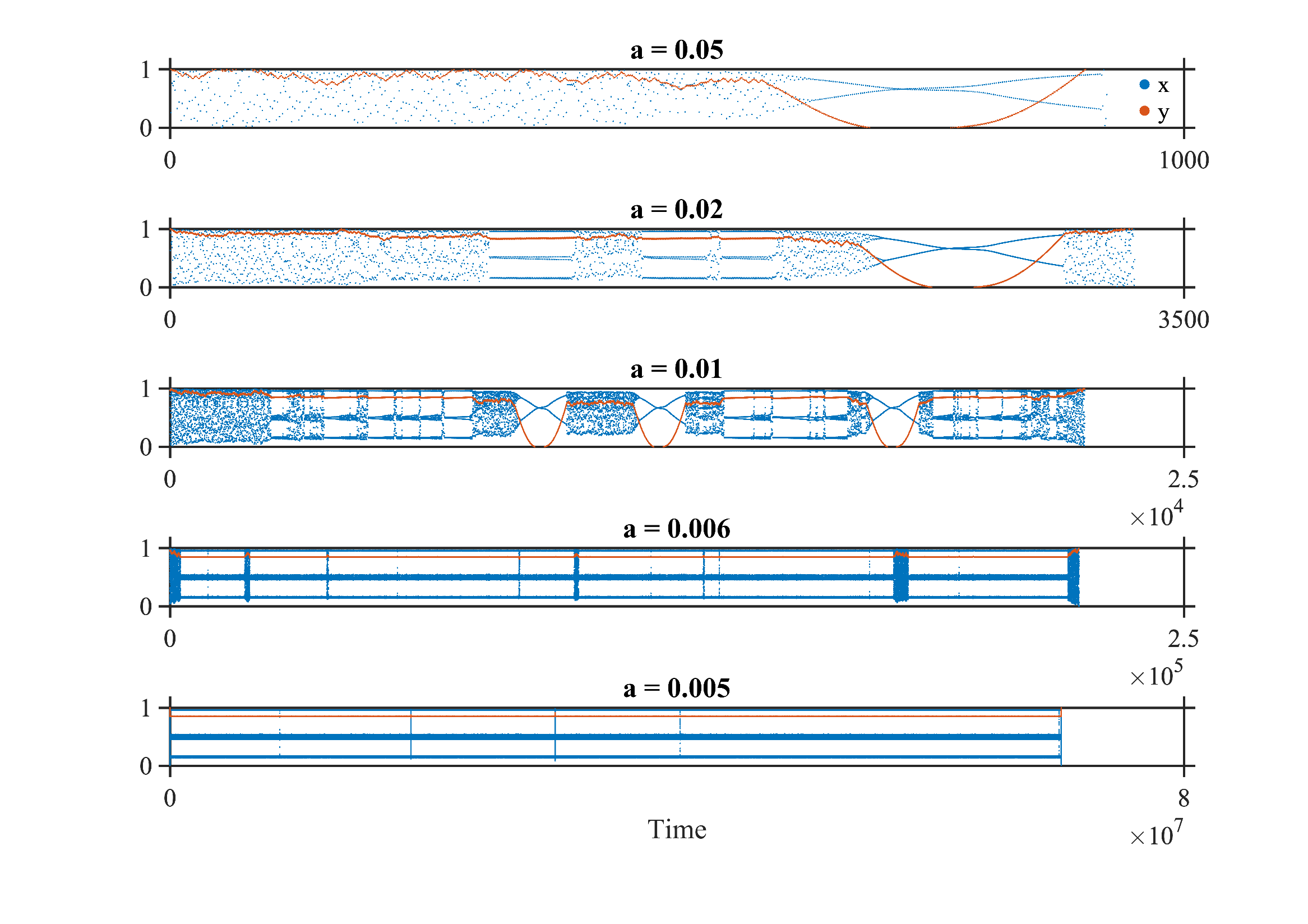}
\caption{\textbf{Time series of $x$ and $y$ with varying $a$.}
For relatively large $a$ ($a = 0.05$ and $a = 0.02$, first two panels), transient attractors associated with periodic windows exhibit weak resilience, resulting in short lifetime within the phase space ($10^3$ and $3.5 \times 10^3$, respectively). As $a$ decreases ($a = 0.01$, $0.006$, and $0.005$, last three panels), the resilience becomes stronger and the corresponding lifetime increases significantly ($2.5 \times 10^4$, $2.5 \times 10^5$, and $8 \times 10^7$, respectively).} 
\label{fig:NLM_xy_t}
\end{figure*}

\section{Numerical Results}
\label{NLM_Numerical_Results}
In the preceding sections, we established mathematically and demonstrated numerically that the $Y$-dynamics exhibits diffusive motion when $x$ evolves in a chaotic regime, whereas in periodic regimes $Y$ undergoes slow monotonic drift, causing the corresponding periodic regions to act as transient attractors. Moreover, the periodic window, in general, is more attractive than the period-doubling bifurcation region.

Now, let us further analyze the dynamics in terms of the parameter $a$. As defined in our model, $a$ plays distinct dynamical roles in the chaotic and periodic regimes. In the chaotic regime, $a$ enters the effective diffusion coefficient $D$ and therefore affects the strength of the diffusive motion. In periodic regions, the cycle-to-cycle displacement in the $Y$-direction is $a\delta_i$, so that $|a\delta_i|$ determines the
drift per cycle and, near the boundary of a periodic region, the corresponding escape step. As discussed in Section~\ref{subsec:periodic_x}, the parameter $a$ therefore scales the effective resilience of the transient attractor. The smaller $a$ is, the greater the basin of resilience: trajectories are more readily retained within the basin of attraction and are less likely to be expelled from it.

The transient time series of $x$ and $y$ are illustrated in \autoref{fig:NLM_xy_t} for different values of the parameter $a$. Starting from a random initial condition $x_0 \in (0, 1)$ with $y_0=1$ and $r=3$, representative trajectories are shown for several values of $a$. Each trajectory is simulated until it escapes through the boundary $R=4$, which in this case corresponds to $y=1$. 

For all five panels, when $x$ evolves chaotically, $y$ exhibits diffusive motion; when $x$ evolves within a periodic window or the period‑doubling bifurcation region, the corresponding periodic region acts as a transient attractor. As $a$ decreases, the resilience of the
transient attractor increases, so that trajectories are retained more effectively within the basin of attraction and remain there for longer times. In particular, as discussed earlier, the periodic window exhibits stronger transient attraction than the period-doubling bifurcation region. As the resilience increases, this stronger attraction becomes increasingly pronounced, leading to the distinct behavior observed among the different panels.

In the first panel, for relatively large $a$ ($a=0.05$), the drift per cycle in the $Y$-direction $|a\delta_i|$ is correspondingly large, resulting in significant drift. This strong drift prevents trajectories from entering the periodic window, which is very narrow in the $Y$-direction. Thus, trajectories predominantly evolves through the chaotic region and the period-doubling bifurcation region. In the second panel, with a reduced value of $a$ ($a=0.02$), Eq.~\eqref{eq:NLM_quasi_condition1} is satisfied only for a narrow set of initial conditions. Furthermore, the basin of resilience remains quite weak. Consequently, although trajectories are temporarily captured by the periodic window, their lifetime within it is short. In the last three panels, $a$ is reduced further to $0.01$, $0.006$, and $0.005$. As Eq.~\eqref{eq:NLM_quasi_condition1} and Eq.~\eqref{eq:NLM_quasi_condition2} are more readily satisfied, trajectories are more effectively retained within the period-$L$ windows and are less likely to be expelled from them. Consequently, the lifetime within periodic windows increases from the order of $10^4$ to the order of $10^7$.

The above results show how the parameter $a$ scales the effective resilience of the transient attractor. We next examine the survival probability to characterize the distinct dynamical roles played by $a$ in diffusive motion and transient attraction. Here, the survival
probability describes the lifetimes of trajectories before they escape through the boundary $R=4$. As discussed before, the dynamics of $y$ can be very different depending on whether the $x$-dynamics being accumulated is chaotic or
periodic. When $x$ evolves chaotically, $y$ exhibits diffusive motion; when $x$ evolves periodically, $y$ undergoes slow monotonic drift,
causing the corresponding periodic regions to act as transient attractors. Correspondingly, the survival probability is directly linked to these distinct dynamical regimes.

As explained in Section~\ref{subsec:chaotic_x}, when $x$ evolves in a chaotic regime, the corresponding $y$-dynamics is well approximated by a Wiener process for sufficiently large $n$. Thus,
\begin{equation}
    \langle Q_n \rangle = 0, \quad \langle Q_n^2 \rangle \sim n,
\end{equation}
where $Q_n=\frac{y_n}{\sqrt{D}}$ and, when the contribution from nonzero-lag correlations is negligible to leading order, $D\approx a^2\sigma^2$. Thus, the first-passage time distribution of a standard Wiener process to a the target point $Q^*$ is given by \cite{2001_Redner_book}:
\begin{equation}
g (t)= \frac{Q^*}{\sqrt{2\pi t^3}}e^{-\frac{(Q^*)^2}{2t}}.
\end{equation}
This equation can be further rewritten as
\begin{equation}
g(t)=
\frac{y^*/\sqrt{D}}{\sqrt{2\pi t^3}}
\exp\left[-\frac{(y^*/\sqrt{D})^2}{2t}\right].
\end{equation}
For large $t$, the exponential factor approaches unity, and the
first-passage-time density exhibits the asymptotic power-law decay
\begin{equation}
g(t)\sim \frac{1}{a}\,t^{-3/2}.
\end{equation}
During the initial regime, the surviving ensemble is dominated by trajectories undergoing chaotic diffusion and not captured by transient attractors. In our system, the onset of escape corresponds to reaching a target at a small positive $y^*>0$. Let $N(t)$ denote the number of surviving trajectories. Since the first-passage-time density determines their loss rate, we have
\begin{equation}
-\frac{dN}{dt}\propto g(t)
\sim \frac{1}{a}\,t^{-3/2}.
\end{equation}
For a fixed initial size, integration therefore gives
\begin{equation}
N(t)\sim \frac{1}{a}\,t^{-1/2},
\label{eq:N_PowerLaw}
\end{equation}
in the diffusion-dominated power-law regime, with exponent $-1/2$ and prefactor proportional to $a^{-1}$. For a finite ensemble, let $t_c$ denote 
the cutoff time at which the power-law regime terminates, i.e., when the number of surviving trajectories contributing to this regime becomes of order unity. Thus, $\frac{1}{a}\,t_c^{-1/2}\sim O(1)$, which gives
\begin{equation}
t_c\sim \frac{1}{a^2}.
\label{eq:NLM_cutoff}
\end{equation}
Hence, decreasing $a$ corresponds to a longer cutoff time $t_c$, thereby extending the duration of the initial power-law regime.

\begin{figure*}[t]
\centering
\renewcommand{\figurename}{Fig.}
\includegraphics[width=\linewidth]{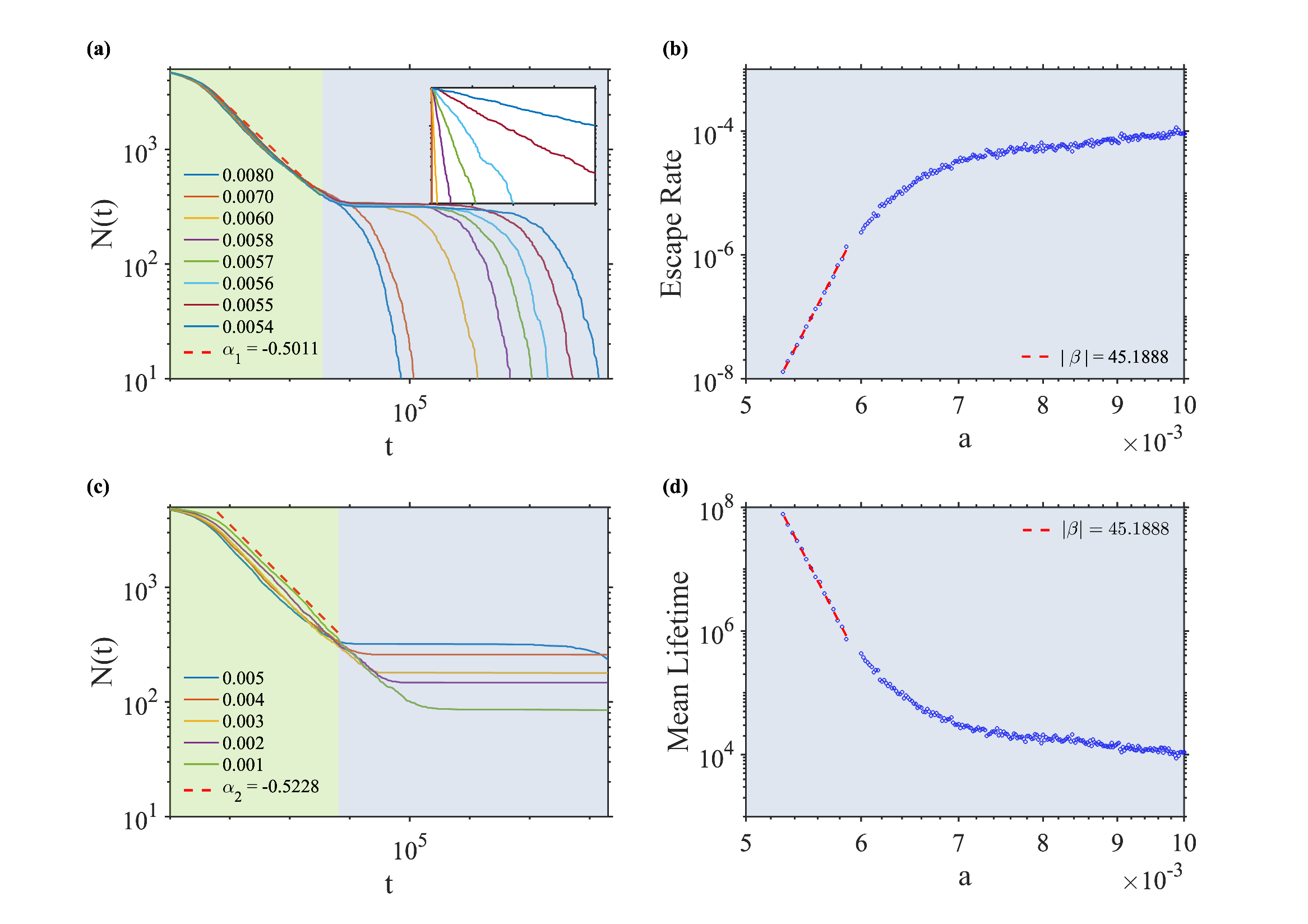} 
\caption{\textbf{Number of survival trajectories $N(t)$ with varying $a$.} Green and blue backgrounds mark two distinct decay: power-law decay with exponent $\alpha_1 = -0.5011$ and $\alpha_2 = -0.5228$ (red dashed lines in (a) and (c)), and exponential decay, confirmed by the linear behavior in the semilogarithmic inset of (a). The corresponding escape rate and mean lifetime for exponential decay are further calculated in (b) and (d) for $a \in (0.005, 0.010)$. By fitting the curve to a power-law form, the escape rate scales as $\sim a^{|\beta|}$, and mean lifetime scales as $\sim \frac{1}{a^{|\beta|}}$, with exponent $|\beta|=45.1888$ (red dashed lines in (b) and (d)).  %with exponent $|\beta| = 8105$ (black dashed lines in (b) and (d)). 
In panel (c), by choosing small but substantially different values of $a$ ($a\in (0.001, 0.005)$), the emergence of a super-long transient tail is observed.} 
\label{fig:NLM_Sur_a}
\end{figure*}

Initialized the system with $N(0)=5000$ and random initial conditions $x_0 \in (0, 1)$, with $y_0=1$ and $r=3$. \autoref{fig:NLM_Sur_a}   (a) and (c) show the corresponding changes of $N(t)$ for different values of $a$. $N(t)$ exhibits a distinct two-stage decay, reflecting the two distinct dynamical regimes identified in
Section~\ref{NLM_System_Dynamics_Analysis}.

First, for trajectories whose $x$-dynamics is predominantly chaotic and are not captured by transient attractors, the trajectories escape from the phase space first, and the number of surviving trajectories $N(t)$ initially exhibits a power-law decay. The corresponding exponents are $\alpha_1 = -0.5011$ and $\alpha_2 = -0.5228$ (red dashed lines in panels (a) and (c), respectively), highlighted by the green-shaded background. The slight deviations of these numerical values with the theoretical exponent $-1/2$ (see Eq.~\eqref{eq:N_PowerLaw}) are attributed to a small group of trajectories governed by periodic behavior. This initial power-law regime persists until the cutoff time $t_c$, which increases as the parameter $a$ decreases. This dynamics is clearly illustrated in panel (c), where $N(t)$, computed for substantially different values of $a$, exhibits markedly different cutoff times. For example, as $a$ decreases from $0.005$ (blue curve) to $0.001$ (green curve), the cutoff time $t_c$ increases by approximately one order of magnitude, from $10^4$ to $10^5$.

Beyond $t_c$, most trajectories dominated by chaotic diffusion have already escaped, while the remaining trajectories are primarily trapped within transient attractors. As these long-lived trapped trajectories gradually escape, the decay of $N(t)$ crosses over from the initial power-law regime to an approximately exponential regime, highlighted by the blue shaded background. A detailed view of this exponential regime is presented in the inset of panel (a). The linear behavior of $N(t)$ in this semilogarithmic plot confirms its exponential behavior. Moreover, as discussed earlier, the parameter $a$ scales the effective resilience of the transient attractor: as $a$ decreases, the resilience increases, allowing trajectories to be retained more effectively within the periodic window and reducing their likelihood of escape. Consequently, trajectories spend longer on the transient attractor. As shown in panel (a), when $a$ decreases from $0.008$ to $0.0054$, the lifetime increases from approximately $10^4$ to $10^8$. To quantify this dependence, the escape rates and corresponding mean lifetimes (defined as the inverse of the escape rates) associated with the exponential decay regime are further calculated for $a \in (0.005, 0.010)$. The results are shown in panels (b) and (d), respectively. As discussed above, decreasing $a$ leads to a longer lifetime within periodic regions and, correspondingly, a lower escape rate. In particular, in the limit $a \rightarrow 0$, the escape rate is expected to approach zero, while the mean lifetime diverges. To characterize this dependence quantitatively, we fit the numerical results in panel (b) using a power-law form. Near $a=0.005$, the escape rate is well described by $\sim a^{|\beta|}$, with exponent $|\beta|=45.1888$ (red dashed line). Accordingly, the corresponding mean lifetime, shown in panel (d), scales as $\sim a^{-|\beta|}$. Thus, as $a \rightarrow 0$, the escape rate tends to zero, whereas the mean lifetime tends to infinity, indicating the emergence of increasingly long transients. Other functional forms may also provide suitable fits to the red curve in panels (b) and (d), as long as they preserve the opposite dependence of the escape rate and mean lifetime on $a$: decreasing $a$ suppresses the escape rate while increasing the mean lifetime. In the limit $a \rightarrow 0$, the escape rate approaches zero, whereas the mean lifetime diverges. In panel (c), by choosing small but substantially different values of $a$ ($a \in (0.001, 0.005)$), the emergence of long transient tails is observed as $a$ decreases. Here, we are unable to obtain the escape rate or lifetime numerically in the limit of $a\to 0$, but we can see that the lifetime grows extremely large as $a \to 0$. These are the long transients that we discuss in this work.

The parameter $a$, as the sole control parameter in our system, plays distinct roles depending on whether $x$ evolves in a chaotic or periodic regime. When $x$ evolves within the chaotic region, the corresponding $Y$ exhibits diffusive motion, and $a$ affects the strength of this motion through the effective diffusion coefficient $D$. As $a$ decreases, the duration of the power law regime increases. In contrast, when $x$ evolves within a periodic window or the period-doubling bifurcation region, the corresponding periodic region acts as a transient attractor. In this case, $a$ scales the effective resilience of the transient attractor through the drift per-cycle $|a\delta_i|$. As $a$ decreases, the escape step becomes smaller and the basin of resilience increases: trajectories are more readily retained within the basin of attraction and are less likely to be expelled from it. Consequently, the lifetime on the transient attractor increases.  Overall, decreasing $a$ increases the duration of both diffusive motion and transient attractors. However, their effects are fundamentally different.

Although the present model is not intended as a realistic population model, it can, like the logistic map itself, be given a population-dynamical interpretation that provides a simple theoretical perspective on the underlying mechanisms.

As May\cite{May_1976} noted for the logistic map, if trajectories exceed unity, subsequent iterations diverge to infinity, which can be interpreted as population extinction. Under this interpretation, the number of surviving trajectories $N(t)$ can also be given a population-dynamical meaning. For example, if different initial conditions $x_0$ are interpreted as independently (without interaction) evolving subpopulations or genetic variants of the same species, then $N(t)$ may be viewed as reflecting the number of surviving subpopulations or variants. Alternatively, if each trajectory is interpreted as representing a distinct species, $N(t)$ may be viewed as reflecting species richness.

Within this theoretical interpretation, decreasing $a$ produces two opposing effects. On the one hand, a smaller $a$ prolongs the diffusive regime, allowing more trajectories to escape before being captured by long-lived transient attractors. Consequently, the number of surviving trajectories $N(t_c)$ decreases, which can be interpreted as a reduction in the number of surviving subpopulations, genetic variants, or species. On the other hand, once trajectories are captured by a periodic transient attractor, a smaller $a$ increases its effective resilience and prolongs their lifetime on the attractor, corresponding to a relatively stable and favorable environment for species persistence.

In this sense, decreasing $a$ has two opposing effects: on the one hand, it reduces the number of surviving trajectories during the diffusive motion, which can be interpreted as a reduction in genetic diversity or species richness; on the other hand, it prolongs the lifetime on the transient attractor, whose quasi-periodic dynamics provide a relatively stable and favorable environment for species persistence.  These opposing effects indicate a trade-off between biodiversity and stability. This suggests that it is not possible to maximize both species richness and system stability simultaneously. There exists a critical value $a_c$ that balances species richness and system stability, thereby optimizing ecosystems. This biodiversity–stability trade-off may contribute to the persistence and success of natural systems. 

Importantly, this interpretation does not imply that the present model quantitatively describes a real ecosystem; rather, it illustrates how competing effects on diversity and persistence can arise intrinsically from the same feedback mechanism.

\section{Generality of the alternating-feedback mechanism}
\label{sec:generality}

The preceding analysis was developed using the logistic map as a minimal example. However, the underlying mechanism does not depend on the specific functional form. More generally, we consider
\begin{align}
    x_{n+1}
    &=
    F(x_n; R_n), 
    \nonumber\\
    y_{n+1}
    &=
    y_n+(-1)^n a(x_n-\ell),
    \label{eq:general_models}
\end{align}
where $F(x;R)$ denotes a one-parameter family of one-dimensional maps, $x \in I$ is the dynamical variable, $R$ is the control parameter, $\ell\in I$ is an arbitrary but fixed constant, and $r$ is a fixed parameter. Importantly, the functional dependence of $F$ on $R$ is left arbitrary. For example, the logistic and quadratic maps correspond respectively to
\begin{equation}
    F(x_n;R_n)=R_nx_n(1-x_n),
    \qquad
    F(x_n;R_n)=R_n-x_n^2,
\end{equation}
and are therefore both contained in Eq.~\eqref{eq:general_models},
despite the different ways in which the control parameter $R_n$ enters the map.

\subsection{Chaotic $x$-Dynamics: Diffusive Motion in $y$}
Suppose $x_n$ evolves in a chaotic regime satisfying sufficiently strong mixing conditions. The $y$-dynamics can be decomposed as
\begin{equation}
    y_n
    =
    y_n^{(\mathrm{bnd})}
    +
    y_n^{(\mathrm{diff})}.
\end{equation}
As shown in \hyperref[box:chaotic_diffusive]{Appendix.~\ref*{box:chaotic_diffusive}}, the term $y_n^{(\mathrm{bnd})}$ remains uniformly bounded for all $n$:
\begin{equation}
    \left|
    y_n^{(\mathrm{bnd})}
    \right|
    \leq
    |a(c-\ell)|
    \label{eq:bounded_term}
\end{equation}
due to the alternating sign $(-1)^n$, and therefore does not affect the diffusive behavior. In contrast, the term $y_n^{(\mathrm{diff})}$ satisfies
\begin{equation}
    \left\langle
    \left(
    y_n^{(\mathrm{diff})}
    \right)^2
    \right\rangle
    \sim
    Dn,
    \qquad
    n\rightarrow\infty,
\end{equation}
where
\begin{equation}
    D
    =
    a^2
    \left[
        C(0)
        +
        2\sum_{k=1}^{\infty}
        (-1)^k C(k)
    \right]
\end{equation}
is the effective diffusion coefficient. Under diffusive rescaling, it converges weakly to a Wiener process
\begin{equation}
    \frac{
        y_{\lfloor Nt\rfloor}^{(\mathrm{diff})}
    }{
        \sqrt{DN}
    }
    \Longrightarrow
    W(t),
    \qquad
    N\rightarrow\infty.
\end{equation}
Since $y_n^{(\mathrm{bnd})}$ is uniformly bounded,
\begin{equation}
    \frac{
        y_{\lfloor Nt\rfloor}^{(\mathrm{bnd})}
    }{
        \sqrt{DN}
    }
    \longrightarrow
    0,
    \qquad
    N\rightarrow\infty.
\end{equation}
Hence, $y_n^{(\mathrm{bnd})}$ is negligible on diffusive scales, thus we have $y$-dynamics satisfies
\begin{equation}
    \frac{
        y_{\lfloor Nt\rfloor}
    }{
        \sqrt{DN}
    }
    \Longrightarrow
    W(t),
    \qquad
    N\rightarrow\infty.
\end{equation}
This indicates that $y$-dynamics exhibits diffusive scaling and, after diffusive rescaling, converges weakly to a Wiener process.

The above derivation reveals two key ingredients that contribute to the emergence of diffusive motion in the $y$-direction: (i) the alternating sign $(-1)^n$ ensures that $y_n^{(\mathrm{bnd})}$ remains uniformly bounded and, therefore, becomes negligible on diffusive scales; and (ii) sufficiently strong chaotic mixing, together with the corresponding functional central limit theorem, causes $y_n^{(\mathrm{diff})}$ to exhibit diffusive scaling and, under diffusive rescaling, to converge weakly to a Wiener process.

Importantly, this derivation does not depend on the particular functional form of $F(x; R)$, but only on the properties of the chaotic dynamics and the alternating-feedback structure. Therefore, the result applies to one-dimensional map whose chaotic regime satisfies sufficiently strong mixing conditions together with the corresponding functional central limit theorem.

%%--------------------------------------------
\subsection{Periodic $x$-Dynamics: Transient Attractor}
Suppose $x_n$ evolves in a period-$L$ window. Owing to the slow drift of $y_n$, as $x_n$ evolves along a quasi-period-$L$ orbit, the corresponding $(X_{i,j},Y_{i,j})$ evolves in a quasi-period-$2L$ orbit.

As shown in \hyperref[box:periodic_ballistic]{Appendix.~\ref*{box:periodic_ballistic}}, the cycle-to-cycle displacement in the $Y$-direction is
\begin{equation}
    Y_{i+1,j}-Y_{i,j}
    =
    a\delta_i,
    \qquad
    \forall\, i,j,
    \label{eq:general_Y_drift}
\end{equation}
where
\begin{equation}
    \delta_i
    :=
    \sum_{j=0}^{2L-1}
    (-1)^jX_{i,j}.
    \label{eq:general_delta_periodic}
\end{equation}
Thus, $|a\delta_i|$ determines the magnitude of the drift between successive quasi-period-$2L$ cycles.

The strength of this drift, however, differs depending on whether $L$ is odd or even. 

For odd $L$, the alternating factor $(-1)^j$ produces pairwise cancellation between corresponding points of the two successive quasi-period-$L$ segments. Since these corresponding points remain close owing to the slow drift in $y$, $\delta_i$ becomes very small. Consequently, for sufficiently small $a$,
\begin{equation}
    |Y_{i+1,j}-Y_{i,j}| \ll 1.
\end{equation}
Thus, the drift in the $Y$-direction is very slow, giving rise to a large number of successive quasi-period-$2L$ orbits within the periodic region and thereby producing strong transient attraction.

For even $L$, by contrast, the corresponding points of the two successive quasi-period-$L$ segments enter $\delta_i$ with the same sign, so this inter-segment cancellation does not occur. Consequently, $|\delta_i|$ is generally larger and trajectories pass through even-period regions more rapidly. Even-period regions can
therefore still act as transient attractors, but their transient attraction is weaker than that of odd-period regions.

The alternating-cancellation mechanism also makes $|S_i|$ small. As shown in
\hyperref[box:periodic_ballistic]{Appendix.~\ref*{box:periodic_ballistic}}, for the subclass of maps in which the control parameter enters multiplicatively,
\begin{equation}
    F(x;R)=Rf(x),
\end{equation}
where $f:I\rightarrow\mathbb{R}$ is an arbitrary nonlinear function independent of $R$, the cycle-to-cycle evolution of $\delta_i$ satisfies
\begin{equation}
    \delta_{i+1}
    \approx
    \delta_i(1+aS_i).
\end{equation}
For sufficiently small $a$, $|S_i|\ll1$ ensures
\begin{equation}
    1+aS_i>0,
\end{equation}
and therefore
\begin{equation}
    \operatorname{sgn}(\delta_i)
    =
    \operatorname{sgn}(\delta_1),
    \qquad
    i\geq1.
\end{equation}
Consequently, the $Y$-dynamics exhibits a monotonic drift whose direction is determined by the initial sign of $\delta_1$.

The above derivation shows that, for sufficiently small $a$, the alternating-feedback mechanism produces monotonic drift in the \(Y\)-direction within the periodic region, with the direction determined by the initial sign of \(\delta_1\).The strength of this drift, however, depends on the parity of $L$, with odd-$L$ period regions exhibiting stronger transient attraction than even-$L$ period regions. The particular functional form of $F(x; R)$ affects the detailed changes of the $X$-orbit, and therefore the precise values of $\delta_i$ and $S_i$, but does not alter the underlying alternating-feedback mechanism. 

The difference between successive quasi-period-$2L$ cycles in the $X$-direction, in contrast, depends on the particular functional form of the underlying map $F(x; R)$. The change between successive cycles in the $x$-direction can be written as
\begin{equation}
    X_{i+1,j}
    =
    X_{i,j}
    +
    a\delta_i f(X_{i+1,j-1}).
\end{equation}
This relation is therefore not specific to the logistic choice $f(x)=x(1-x)$, but applies to any one-dimensional map that can be written in the form $F(x; R)=Rf(x)$.  

For maps with a different functional dependence on $R$, the change between successive cycles in the $X$-direction takes a different explicit form. Such differences modify the detailed updates of the $X$-orbit, and consequently the precise values of $\delta_i$ and $S_i$, but do not alter the underlying alternating-feedback mechanism in the $X$ and $Y$ directions.

%\hyperref[tab:all_systems]{Table.~\ref*{tab:all_systems}}
%In summary, the generality of the mechanism does not rely on a particular functional form of $F(x; R)$ or on a particular way in which the control parameter $R$ enters the map. Instead, it relies on the alternating-feedback structure together with the dynamical properties of the underlying one-dimensional system. When the $x$-dynamics is sufficiently strong chaotic mixing, the accumulated $y$-dynamics exhibits diffusive scaling and converges, under diffusive rescaling, to a Wiener process. When the $x$-dynamics is periodic, the alternating feedback produces pairwise cancellation between corresponding points of successive cycles, resulting in slow monotonic drift in $y$ and thereby turning periodic regions into transient attractors. Different choices of $F(x;R)$ modify the quantitative details of the dynamics, but not this underlying mechanism.

In summary, the generality of the mechanism does not rely on a particular functional form of $F(x; R)$ or on a particular way in which the control parameter $R$ enters the map. Instead, it relies on the alternating-feedback structure together with the dynamical properties of the underlying one-dimensional system. When $x_j$ evolves in a chaotic regime, the alternating sign ensures that $y_n^{(\mathrm{bnd})}$ remains bounded and therefore becomes negligible on diffusive scales. Under sufficiently strong chaotic mixing and the corresponding functional central limit theorem, the term $y_n^{(\mathrm{diff})}$ exhibits diffusive scaling and, after diffusive rescaling, converges weakly to a Wiener process. When $x_j$ evolves in a periodic regime, the alternating-feedback mechanism produces monotonic drift in \(Y\) for sufficiently small \(a\), with the drift becoming particularly slow in odd-period regions owing to the strong pairwise cancellation in \(\delta_i\).

Different choices of $F(x;R)$ modify the quantitative details of the dynamics, but not this underlying mechanism. Thus, depending on whether the $x_j$ being accumulated is chaotic or periodic, the $y$-dynamics exhibits two qualitatively different behaviors: diffusive motion when $x_j$ evolves in a chaotic regime, and monotonic drift in the $Y$-direction when $x_j$ evolves in a periodic regime, in which case the periodic region acts as a transient attractor. We further verify this generality numerically across different one-dimensional maps, observing qualitatively similar dynamical behavior near periodic windows (Supplementary Fig. 1).

%The examples summarized in \hyperref[tab:all_systems]{Table.~\ref*{tab:all_systems}} illustrate this mechanism across a broad range of one-dimensional nonlinear systems. 

%%------------------CONCLUSION---------------
\section{Conclusion}
\label{NLM_Conclusion}

%Equilibrium ecology provides an oversimplified view of natural systems by modeling them as deterministic processes that always converge to a single, stable state, with stochasticity treated merely as an external perturbation\cite{1973_Egerton, 1995_Wu_balance, 1920_lotka1, 1920_Lotka2, Volterra_1926}. In response, a nonequilibrium perspective has emerged, emphasizing that nature is governed by an interplay between deterministic and stochastic processes\cite{2001_Scheffer, 2003_Scheffer, 2009_Scheffer, 1997_Higgins_stochastic}. Importantly, stochasticity---acting as an internal driver---plays an important role in natural systems. The key properties of realistic systems---including the existence of alternative stable states, adaptive capacity, critical transitions, system resilience, and stochasticity---are summarized in Property~\ref{prop:nonequi_ecology}. Currently, a central challenge in nonequilibrium ecology is that no single model can unify all of these dynamics, particularly when stochasticity acts as an internal driver.
In this work, we have developed a minimal alternating-feedback framework that unifies deterministic and stochastic dynamics within a single system. Rather than introducing stochasticity as an external term, we allow stochastic-like dynamics to emerge endogenously from deterministic chaos and to feed back on the deterministic dynamics. This framework allows us to address three central questions.

In our system, we provide a systematic analysis and a mathematically rigorous proof (in \hyperref[box:chaotic_diffusive]{Appendix.~\ref*{box:chaotic_diffusive}} and \hyperref[box:periodic_ballistic]{Appendix.~\ref*{box:periodic_ballistic}}) to help us understand the relationship between deterministic behavior and stochastic processes.
%\paragraph{How do stochastic dynamics emerge intrinsically from deterministic dynamics?} Importantly, this result does not depend on the particular functional form of the underlying map but reflects a general principle of the alternating-feedback structure. Thus, stochasticity is not merely an external perturbation but an intrinsic dynamical outcome that can fundamentally shape the system's evolution.

\paragraph{What role does stochasticity play in long transients?}  

As explained, when $x$ lies within a periodic window, $Y$ monotonically approaches the boundaries of the periodic window. Once $Y_{i, j}$ move slightly outside of the boundary, the near cancellation in $\delta_i$ breaks down, leading to rapid growth of $|\delta_i|$. Consequently, the escape step ($|a\delta_i|$) pushes the trajectory irreversibly out of the periodic window and into the neighboring chaotic region. There, the diffusive motion in the $Y$-direction provides $Y_{i,j}$ with opportunities to re-enter the periodic window, producing repeated cycles of transient trapping in the period-$L$ window. 

Thus, on the one hand, stochasticity is essential for maintaining the persistence of periodic cycles; in other words, \textbf{stochasticity stabilizes the system}. This is further quantified by parameter $a$: when $a$ is small, the escape step is tiny, so trajectories remain close to the basin boundary, increasing the likelihood of re-entering the attracting region through diffusive motion. On the other hand, once $Y_{i,j}$ enters the chaotic region, it may also escape to infinity via the boundary $R=4$. From this perspective, stochasticity contributes to the eventual collapse of the system; that is, \textbf{stochasticity destabilizes the system.}

In summary, stochasticity plays a dual role in our system: on the one hand, it contributes to the persistence of periodic cycles; on the other hand, it facilitates the eventual collapse of the system. Thus, stochasticity acts as both a stabilizing and destabilizing force, dynamically regulating the stability of the system.

\paragraph{What contributes to the emergence of long transients?} 
%Currently, papers about this question...
In our system, when $a$ is small, the corresponding escape step ($|a\delta_i|$) is also small. As a result, trajectories remain close to the basin boundary and are more readily attracted to the transient attractor, producing repeated cycles within the period-$L$ window and thereby contributing to the emergence of super-long transients. This indicates that the emergence of super-long transients requires both $|a|$ and $|\delta_i|$ to be sufficiently small. 
 
The parameter $a$ acts as a coupling coefficient that determines the influence of changes in $x$ on $y$. Specifically, $y_{n+1}$ is updated based on the difference $x_n-x_{n-1}$, scaled by $a$. When $a$ is small, the effective scale between $x$ and $y$ becomes large. This indicates that the \textbf{two variables evolve on different timescales, with one dominating the dynamics while the other remains comparatively small}, as in slow–fast systems. This condition gives rise to the dynamics associated with small $a$.

$\delta_i=\sum_{j=0}^{2L-1}(-1)^jX_{i,j}$ can be sufficiently small only when $X_{i,j}$ lies within an odd-period-$L$ window and the alternating sign $(-1)^j$ induces strong cancellation among $X_{i,j}$ in the sum. In our model, this alternating sign creates a symmetry in the $X$-direction, leading to cancellation for the sum of $X$. In realistic systems, such symmetry in the $X$-direction does not exist---there is no negative population size, for instance. However, the cancellation can be generated through time delay. By summarizing the $y_{n+1}$ and $y_n$ equations in our model, the system can be rewritten as 
\begin{align}
    y_{n+1} &=y_{n-1}+a(x_n-x_{n-1}).
\end{align}
As seen in the $y$-equation, the alternating sign in our equation can be reformulated as \textbf{a time-delay equation, which effectively induces a cancellation in the sum of $X$-direction}---a condition that contributes to the dynamics associated with small $\delta$.

In summary, two factors contribute to the emergence of long transients: (i) the presence of variables evolving on widely separated timescales, and (ii) time-delay–induced cancellation in the sum of certain variables.

\paragraph{How do stochastic and deterministic dynamics interact to produce complex behavior?} In our system, chaotic $x$-dynamics generates diffusive motion in the $y$-direction, while periodic $x$-dynamics produces monotonic drift in the $y$-direction within periodic regions, causing them to act as transient attractors. Stochastic dynamics, emerging endogenously from deterministic chaos, and deterministic dynamics are thus unified within a single framework: trajectories drift repeatedly between these two regimes. During this process, there are two types of critical transitions: (i) shifts between alternative stable states, represented by different transient attractors, and (ii) collapse transitions, in which trajectories irreversibly escape from the phase space. The resilience is quantified by how readily trajectories are retained within or expelled from the transient attractors through their interaction with stochastic dynamics. Meanwhile, the feedback variable $y$ continuously adjusts in response to changes in $x$, providing the system with an intrinsic adaptive capacity. These behaviors therefore arise as different features of the same closed feedback loop, linking alternative stable states, adaptive capacity, critical transitions, stochasticity, and resilience within a single dynamical framework. This provides a minimal dynamical route by which multiple characteristic features can coexist and interact within a single model but not independent ingredients. 

This unification suggests that the complexity observed in natural systems—from ecosystems to climate to social systems—may arise not from the multiplicity of external factors but from the intrinsic dynamics of feedback between variables. The interplay between deterministic dynamics and stochastic processes, mediated by endogenous feedback, lies at the heart of this complexity. This perspective is consistent with the view of ecosystems as complex adaptive systems, in which macroscopic organization can emerge from nonlinear local interactions and self-organization \cite{Levin1998}.

\paragraph{A biodiversity--stability trade-off.}
In our framework, the parameter $a$ governs the duration of both the diffusive regime and the transient-attractor regime: decreasing $a$ prolongs both. Under the ecological interpretation, however, these two regimes have opposite effects. A longer diffusive regime allows more trajectories to escape before reaching the transient attractor, thereby reducing the number of surviving trajectories $N(t_c)$, which corresponds to lower genetic diversity or species richness under the ecological interpretation. In contrast, the transient attractor corresponds to the periodic region, which represents a relatively persistent and long-lived dynamical state. A longer lifetime on the attractor therefore indicates prolonged persistence of the surviving trajectories, corresponding to enhanced system stability or resilience. In this sense, decreasing $a$ reduces diversity while increasing stability, giving rise to a biodiversity--stability trade-off: it is not possible to maximize both simultaneously, and there exists a critical value $a_c$ that balances species richness and system stability.

Importantly, the mechanism identified here is not specific to the logistic map used as a minimal example. The underlying alternating-feedback mechanism does not depend on the particular functional form of $F(x;R)$ but extends to a broader class of one-dimensional dynamical systems. This generality suggests that the mechanism reflects fundamental dynamical principles rather than model-specific artifacts.

%%------------------------
%\iffalse
\section*{Appendix}

\subsection{CHAOTIC $x$-DYNAMICS: DIFFUSIVE MOTION in $y$}  
\label{box:chaotic_diffusive}

Suppose $x_n$ evolves in a chaotic regime of an underlying one-dimensional map with state space $I\subseteq\mathbb{R}$. Let $\mu$ denote an invariant probability measure of the map. We restrict attention to an ergodic chaotic regime and assume that the centered observable introduced below satisfies sufficiently strong mixing conditions.

\noindent For an observable $F:I\to\mathbb{R}$, its invariant average is defined as
\begin{equation}
    \langle F(x)\rangle
    :=
    \int_I F(x)\,d\mu(x).
\end{equation}
Since $\mu$ is invariant under the chaotic dynamics, the distribution of $x_n$ with respect to $\mu$ is stationary. In particular,
\begin{equation}
    c
    :=
    \langle x_n\rangle_{\mu}
    =
    \int_I x\,d\mu(x),
\end{equation}
which is independent of $n$. We then introduce the centered variable
\begin{equation}
    z_n:=x_n-c.
\end{equation}
Then
\begin{equation}
    \langle z_n\rangle=0,
    \qquad
    \langle z_n^2\rangle=\sigma^2,
\end{equation}
where
\begin{equation}
    \sigma^2
    :=
    \int_I (x-c)^2\,d\mu(x).
\end{equation}
Since the $y$-dynamics is given by
\begin{equation} 
y_n =a\sum_{j=0}^{n-1}(-1)^j(x_j-\ell), 
\end{equation}
where $\ell\in I$ is an arbitrary constant, substituting $x_j=c+z_j$ yields
\begin{align}
    y_n
    &=
    a\sum_{j=0}^{n-1}
    (-1)^j(c+z_j-\ell)
    \nonumber\\
    &=
    a(c-\ell)
    \sum_{j=0}^{n-1}(-1)^j
    +
    a\sum_{j=0}^{n-1}(-1)^jz_j
    \nonumber\\
    &=
    y_n^{(\mathrm{bnd})}
    +
    y_n^{(\mathrm{diff})}.
\end{align}
The bounded term is
\begin{equation}
    y_n^{(\mathrm{bnd})}
    =
    a(c-\ell)\sum_{j=0}^{n-1}(-1)^j
    =
    \begin{cases}
        a(c-\ell), & n\ \text{odd},\\[3pt]
        0, & n\ \text{even},
    \end{cases}
\end{equation}
which remains uniformly bounded for all $n$:
\begin{equation}
    \left|
    y_n^{(\mathrm{bnd})}
    \right|
    \leq
    |a(c-\ell)|.
    \label{eq:bounded_term}
\end{equation}
Thus, the contribution associated with the arbitrary constant $\ell$ remains bounded due to the alternating sign and does not affect the
asymptotic diffusive behavior.\\

\noindent Let
\begin{equation}
    \eta_j:=(-1)^jz_j.
\end{equation}
Then the diffusive term can be written as
\begin{equation}
    y_n^{(\mathrm{diff})}
    =
    a\sum_{j=0}^{n-1}\eta_j.
\end{equation}
Since $\langle z_j\rangle=0$, we have
\begin{equation}
    \langle \eta_j\rangle=0,
\end{equation}
and consequently
\begin{equation}
    \left\langle
    y_n^{(\mathrm{diff})}
    \right\rangle
    =0.
\end{equation}

\noindent Let
\begin{equation}
    C(k)
    :=
    \langle z_nz_{n+k}\rangle
\end{equation}
denote the correlation function of the centered chaotic process. The correlation function of the alternating process $\eta_n$ is
\begin{align}
    \widetilde C(k)
    &:=
    \langle\eta_n\eta_{n+k}\rangle
    \nonumber\\
    &=
    \left\langle
    (-1)^nz_n
    (-1)^{n+k}z_{n+k}
    \right\rangle
    \nonumber\\
    &=
    (-1)^k
    \langle z_nz_{n+k}\rangle
    \nonumber\\
    &=
    (-1)^kC(k).
\end{align}
The variance of the diffusive term is therefore
\begin{align}
    \left\langle
    \left(
    y_n^{(\mathrm{diff})}
    \right)^2
    \right\rangle
    &=
    a^2
    \left\langle
    \left(
    \sum_{j=0}^{n-1}\eta_j
    \right)^2
    \right\rangle
    \nonumber\\
    &=
    a^2
    \sum_{j=0}^{n-1}
    \sum_{k=0}^{n-1}
    \langle\eta_j\eta_k\rangle
    \nonumber\\
    &=
    a^2
    \sum_{j=0}^{n-1}
    \sum_{k=0}^{n-1}
    \widetilde C(|k-j|)
    \nonumber\\
    &=
    a^2
    \left[
        n\widetilde C(0)
        +
        2\sum_{k=1}^{n-1}
        (n-k)\widetilde C(k)
    \right]
    \nonumber\\
    &=
    a^2
    \left[
        \widetilde C(0)
        +
        2\sum_{k=1}^{n-1}
        \left(
        1-\frac{k}{n}
        \right)
        \widetilde C(k)
    \right]n. 
    \label{eq:y_fluc_middle}
\end{align}
Under the sufficiently strong chaotic mixing conditions considered here, the correlations are absolutely summable\cite{},
\begin{equation}
    \sum_{k=1}^{\infty}|C(k)|<\infty.
    \label{eq:absolute_summability}
\end{equation}
Since $|\widetilde C(k)|=|C(k)|$,
Eq.~\eqref{eq:absolute_summability} also implies
\begin{equation}
    \sum_{k=1}^{\infty}
    |\widetilde C(k)|<\infty.
\end{equation}
Hence, as $n\to\infty$,
\begin{equation}
    \sum_{k=1}^{n-1}
    \left(1-\frac{k}{n}\right)
    \widetilde C(k)
    \longrightarrow
    \sum_{k=1}^{\infty}\widetilde C(k).
\end{equation}
Consequently, Eq.~\eqref{eq:y_fluc_middle} can be written asymptotically as
\begin{align}
    \left\langle
    \left(
    y_n^{(\mathrm{diff})}
    \right)^2
    \right\rangle
    &\sim
    a^2
    \left[
        \widetilde C(0)
        +
        2\sum_{k=1}^{\infty}
        \widetilde C(k)
    \right]n
    \nonumber\\
    &=
    a^2
    \left[
        C(0)
        +
        2\sum_{k=1}^{\infty}
        (-1)^k C(k)
    \right]n,
    \qquad n\to\infty.
\end{align}
The corresponding effective diffusion coefficient is therefore
\begin{equation}
    D
    :=
    a^2
    \left[
        C(0)
        +
        2\sum_{k=1}^{\infty}
        (-1)^kC(k)
    \right],
    \label{eq:effective_diffusion}
\end{equation}
where $C(0)=\sigma^2$. Thus, we obtain
\begin{equation}
    \left\langle
    \left(
    y_n^{(\mathrm{diff})}
    \right)^2
    \right\rangle
    \sim
    Dn,
    \qquad
    n\rightarrow\infty.
    \label{eq:diffusive_scaling_general}
\end{equation}
For the strongly mixing chaotic regimes considered here, the nonzero-lag correlations decay rapidly and contribute only subleading corrections to the correlation sum. Thus,
\begin{equation}
    \left|
    2\sum_{k=1}^{\infty}
    (-1)^kC(k)
    \right|
    \ll
    \sigma^2,
\end{equation}
so that the contribution from nonzero lags is negligible to leading order, and the effective diffusion coefficient is approximated by
\begin{equation}
    D\approx a^2\sigma^2.
    \label{eq:D_approx_sigma}
\end{equation}
Rescaling 
\begin{equation}
    Q_n :=\frac{y_n^{(\mathrm{diff})}}{\sqrt{D}},
\end{equation}
we obtain 
\begin{equation}
    \langle Q_n \rangle =0, \quad \langle Q_n^2 \rangle \sim n.
\end{equation}
Thus, $Q_n$ exhibits the characteristic diffusive scaling of a Wiener process, with the iteration index $n$ playing the role of discrete time. \\

\noindent To formulate this limiting behavior more precisely, introduce a large time-scale parameter $N$ and rescale the discrete iteration index $n$ according to $t=n/N$. The corresponding rescaled process is defined by
\begin{equation}
Q^{(N)}(t)
:=
\frac{1}{\sqrt{N}}
Q_{\lfloor Nt\rfloor}
=
\frac{
y_{\lfloor Nt\rfloor}^{(\mathrm{diff})}
}{\sqrt{DN}},
\qquad t\geq 0.
\end{equation}
Under sufficiently strong chaotic mixing conditions, the functional central limit theorem applies, yielding
\begin{equation}
Q^{(N)}(\cdot)
\Rightarrow
W(\cdot),
\qquad N\to\infty,
\end{equation}
where $W(t)$ denotes a standard Wiener process. \\

\noindent The bounded term $y_n^{\mathrm{(bnd)}}$ is bounded, whereas the diffusive term grows as 
\begin{equation}
    |y_n^{(\mathrm{diff})}| \sim \sqrt{\langle (y_n^{(\mathrm{diff})})^2 \rangle} \sim \sqrt{Dn}.
\end{equation}
Consequently, 
\begin{equation}
    \frac{|y_n^{(\mathrm{bnd})}|}{|y_n^{(\mathrm{diff})}|}=O(\frac{1}{\sqrt{n}}) \rightarrow 0 \quad (n \rightarrow \infty).
\end{equation}
Thus, the bounded term is negligible for large $n$. The $y_n$ exhibits diffusive scaling, and it behaves as a Wiener process after normalization.

%%------------------------
\subsection{PERIODIC $x$-DYNAMICS: TRANSIENT ATTRACTOR}
\label{box:periodic_ballistic}

%Suppose $x_j$ lies within a period-$L$ window. As an illustrative example, we consider the well-known period-3 window; that is, $R_n=r+y_n$ lies in the interval $(1+\sqrt{8}, 3.857)$\cite{Strogatz_2015}.

Suppose $x_j$ evolves in a period-$L$ window of the one-parameter family $F(x;R)$. Since the $y$-dynamics evolves as
\begin{equation}
    y_n
    =
    y_0
    +
    a\sum_{j=0}^{n-1}(-1)^j(x_j-\ell),
    \label{eq:appendix_y_sum}
\end{equation}
a strictly period-$L$ orbit for $x_n$ would force $y_n$ into a strictly period-$2L$ orbit. However, the slow drift of $y_n$ on $x_n$ in Eq.~\eqref{eq:map1} prevents $x_n$ from being strictly periodic. Instead, $x_n$ exhibits a quasi‑period‑$L$ orbit, while $y_n$ exhibits a corresponding quasi‑period‑$2L$ orbit.

\noindent To describe the slow drift between successive quasi-cycles, we introduce a double index notation: let $(X_{i,j}, Y_{i,j})$ denote the $j$-th point ($j=0, 1, \ldots, 2L-1$) of the $i$-th ($i=0,1, \ldots$) quasi-cycle. The first quasi-period-$L$ orbit is
\begin{equation}
    X_{0,0},X_{0,1},\ldots,X_{0,L-1}.
\end{equation}
Owing to the slow drift in the $y$-direction, we assume that the second quasi-period-$L$ orbit differs only slightly from the first one:
\begin{equation}
    X_{0,L+k}
    =
    X_{0,k}+\epsilon_k,
    \qquad
    k=0,1,\ldots,L-1,
    \label{eq:quasi_cycle_close}
\end{equation}
with $|\epsilon_k|\ll1$. These points constitute the first quasi-period-$2L$ orbit ($i=0$):
\begin{equation}
    (X_{0,0}, Y_{0,0}),\; (X_{0,1}, Y_{0,1}), \; \ldots,\; (X_{0,2L-1}, Y_{0,2L-1}).
\end{equation}

\noindent The first point ($j=0$) of the second period-$2L$ orbit ($i=1$) can be expressed as
\begin{align}
    X_{1,0} 
    &= X_{0,0}+\epsilon^\prime, \nonumber \\
    Y_{1,0}
    &=Y_{0,0}+a\sum_{j=0}^{2L-1}(-1)^j(X_{0,j}-\ell) \nonumber \\
    &=Y_{0,0}+a\sum_{j=0}^{2L-1}(-1)^jX_{0,j} \nonumber \\
    &=Y_{0,0}+a\delta_0,
\end{align}
where 
\begin{equation}
 \delta_i
 =\sum_{j=0}^{2L-1}(-1)^jX_{i,j},
\end{equation}
and the contribution from the arbitrary constant $\ell$ cancels pairwise over the $2L$ successive updates.  Then, according to the general model
\begin{equation}
    x_{n+1}=F(x_n;R_n),
    \qquad
    R_n=r+y_n,
\end{equation}
the next point ($j=1$) of the second quasi-cycle ($i=1$) satisfies
\begin{align}
    X_{1,1}
    &=
    F(X_{1,0};r+Y_{1,0})
    \nonumber\\
    &=
    F(X_{1,0};r+Y_{0,0}+a\delta_0),
    \label{eq:X11_general}
    \\
    Y_{1,1}
    &=
    Y_{1,0}+a(X_{1,0}-\ell)
    \nonumber\\
     & \quad (X_{1,0}\approx X_{0,0} \; \text{when} \; |\epsilon^\prime| \ll 1) \nonumber \\
    &\approx
    Y_{0,0}+a\delta_0+a(X_{0,0}-\ell)
    \nonumber\\
    &=
    Y_{0,1}+a\delta_0.
    \label{eq:Y11_general}
\end{align}
Here, the approximation is based on the assumption that
$X_{1,0}\approx X_{0,0}$ with $|\epsilon'|\ll1$. Using the same approach, we obtain
\begin{align}
    X_{i+1,j}
    &\approx
    F\!\left(
    X_{i+1,j-1};
    r+Y_{i,j-1}+a\delta_i
    \right),
    \nonumber\\
    Y_{i+1,j}
    &\approx
    Y_{i,j}+a\delta_i,
    \qquad \forall\,i,j.
    \label{eq:general_cycle_update}
\end{align}
The explicit change in the $X$-direction therefore depends on the particular functional form of $F(x;R)$, whereas the cycle-to-cycle change in the $Y$-direction is controlled by the common quantity $a\delta_i$. For the subclass of maps in which the control parameter enters multiplicatively,
\begin{equation}
    F(x;R)=Rf(x),
    \label{eq:multiplicative_subclass}
\end{equation}
where $f:I\rightarrow\mathbb{R}$ is independent of $R$, $X_{1,1}$ can be rewritten as 
\begin{align}
    X_{1,1} 
    &=(r+Y_{1,0})f(X_{1,0}) \nonumber \\
    &= (r+Y_{0,0}+a\delta_0)f(X_{1,0}) \nonumber \\
    &=(r+Y_{0,0})f(X_{1,0})+a\delta_0 f(X_{1,0})\nonumber \\
    & \quad (X_{1,0}\approx X_{0,0} \; \text{when} \; |\epsilon^\prime| \ll 1) \nonumber \\
    & \approx (r+Y_{0, 0})f(X_{0,0}) + a\delta_0 f(X_{1,0}) \nonumber \\
    &=X_{0,1}+a\delta_0f(X_{1,0}), 
\end{align}
For this multiplicative subclass, using the same approach, Eq.~\eqref{eq:general_cycle_update} can be written explicitly as
\begin{align}
    X_{i+1,j} &\approx X_{i,j} + a\delta_i f(X_{i+1,j-1}), && (i, j) \neq (0, 0), \nonumber \\
    Y_{i+1,j} &\approx Y_{i,j} + a\delta_i, && (\forall\; i, j).
    \label{box:eq:periodicity}
\end{align}
For $j=0$, the orbit index is understood cyclically, with
\begin{equation}
    X_{i+1, -1} :=X_{i,2L-1}.
\end{equation}
%Thus, although the detailed evolution of $x$ is map dependent, the alternating-cancellation mechanism governing the slow evolution of $y$ remains unchanged.
%This equation can be applied directly to the maps in Table~\ref{tab:all_systems} whose control parameter enters multiplicatively in the form $F(x; R)=R_nf(x_n)$. The quadratic and Pomeau--Manneville maps have a different dependence on $R_n$ and therefore lead to a different explicit expression for the $x$-equation. Nevertheless, applying the same approach to these maps yields relations analogous to Eq.~\eqref{box:eq:periodicity} as well. 

\noindent Eq.~\eqref{box:eq:periodicity} are derived under the approximation that the corresponding points of two successive quasi-cycles remain close, $X_{i+1, j} \approx X_{i,j}$, which can be considered in two cases:
\begin{equation}
\begin{cases}
X_{1,0}\approx X_{0,0},
& |\epsilon'|\ll1,
\qquad (i,j)=(0,0),\\[1mm]
X_{i+1,j}\approx X_{i,j},
& |a\delta_i|\ll1,
\qquad (i,j)\neq(0,0).
\end{cases}
\end{equation}

\noindent Accordingly, $\delta_{i+1}$ can be written as
\begin{align}
    \delta_{i+1} 
    &= \sum_{j=0}^{2L-1}(-1)^jX_{i+1,j} \nonumber\\
    &\approx \sum_{j=0}^{2L-1} (-1)^j(X_{i, j}+a\delta_{i}f(X_{i+1, j-1})) \nonumber\\
    &=\sum_{j=0}^{2L-1}(-1)^jX_{i, j}+a\delta_{i}\sum_{j=0}^{2L-1}(-1)^j f(X_{i+1, j-1}) \nonumber \\
    &=\delta_{i}+a\delta_{i}\sum_{j=0}^{2L-1}(-1)^j f(X_{i+1, j-1}) \nonumber \\
    &=\delta_{i}\left [1+a\sum_{j=0}^{2L-1}(-1)^j f(X_{i+1, j-1})\right] \nonumber \\
    &= \delta_i ( 1+aS_i) \quad (i\geq 1), 
\end{align}
where 
\begin{equation}
S_i
=
\sum_{j=0}^{2L-1}(-1)^j
f(X_{i+1,j-1}).
\end{equation}
The sign of $\delta_i$ is preserved across successive quasi-cycles, only when 
\begin{equation}
    1+aS_i>0.
\end{equation}
A sufficient condition is
\begin{equation}
    |aS_i|<1.
\end{equation}
The sign preservation of $\delta_i$ is thus ensured when $|a|$ and $|S_i|$ are sufficiently small. 

\noindent For odd $L$, $S_i$ can be further rewritten by pairing the corresponding terms in the two successive quasi-period-$L$ segments:
\begin{align}
S_i
&=
\sum_{j=0}^{L-1}
(-1)^j
\left[
f(X_{i+1,j-1})
-
f(X_{i+1,L+j-1})
\right],
\label{eq:Si_pairwise}
\end{align}
since
\begin{equation}
    (-1)^{L+j}=-(-1)^{j}.
\end{equation}
Then, for sufficiently small $a$, since corresponding points of the two successive quasi-period-$L$ segments remain close, we have 
\begin{equation}
f(X_{i+1,j-1})
\approx
f(X_{i+1,L+j-1}).
\end{equation}
Therefore, the pairwise differences in Eq.~\eqref{eq:Si_pairwise} are small, yielding
\begin{equation}
|S_i|\ll1.
\end{equation}
Hence, for sufficiently small $a$, 
\begin{equation}
1+aS_i>0.
\end{equation}

\noindent For even $L$, the cancellation has a different origin.
Since
\begin{equation}
    (-1)^{L+j}=(-1)^j,
\end{equation}
the corresponding terms in the two successive quasi-period-$L$ segments enter $S_i$ with the same sign and therefore do not cancel between the two segments. Instead, the cancellation occurs within each quasi-period-$L$ segment through the alternating signs of
successive terms. Using the cyclic indexing, $S_i$ can be grouped as
\begin{align}
S_i
&=
\sum_{j=0}^{L-1}
\left[
f(X_{i+1,2j+1})
-
f(X_{i+1,2j})
\right].
\label{eq:Si_even}
\end{align}
When successive terms within each quasi-period-$L$
segment remain closely matched,
\begin{equation}
    f(X_{i+1,2j+1})
    \approx
    f(X_{i+1,2j}),
\end{equation}
the pairwise differences in Eq.~\eqref{eq:Si_even} are small,
yielding
\begin{equation}
    |S_i|\ll1.
\end{equation}
Hence, for sufficiently small $a$,
\begin{equation}
    1+aS_i>0.
\end{equation}
Above all, for odd \(L\), the cancellation occurs between the two successive quasi-period-\(L\) segments, whereas for even \(L\), it occurs between successive terms within each segment.

\noindent Thus, the sign of $\delta_i$ is preserved across successive quasi-cycles. In particular, for all $i\geq1$,
\begin{equation}
    \operatorname{sgn}(\delta_i)
    =
    \operatorname{sgn}(\delta_1).
\end{equation}
Since 
\begin{equation}
    Y_{i+1,j}-Y_{i,j} \approx a\delta_i,
\end{equation}
we have, for $a>0$:
\begin{equation}
\begin{cases}
    Y_{i+1,j} &>Y_{i,j}, \quad \text{if} \quad \delta_1>0, \nonumber \\
    Y_{i+1,j} &<Y_{i,j}, \quad \text{if} \quad \delta_1<0.
\end{cases}
\end{equation}
Thus, for sufficiently small $a$, $Y_{i,j}$ exhibits a monotonic drift in the $y$-direction across successive quasi-cycles, with the direction of the drift determined by the sign of $\delta_1$. 

\noindent Although the multiplicative subclass $F(x;R)=Rf(x)$ was used above to obtain an explicit expression for the cycle-to-cycle evolution, the underlying alternating-feedback mechanism does not rely on the particular functional form of $F(x;R)$.
%------------------
%\fi
\section*{Acknowledgements}

We are grateful for the stimulating discussions with Christian Beck.

%\section*{Author Contributions}

%Both authors developed the model system together. Misha Chai performed the numerical simulations and created the figures. All authors contributed to the interpretation of the results and to writing the manuscript.

\section*{Competing Interests}
The authors declare that they have no competing financial interests.

\section*{Correspondence}
Correspondence and requests for materials should be addressed to chaimisha@pks.mpg.de.

%Figures and tables can be referenced in LaTeX using the ref command, e.g. Figure \ref{fig:stream} 
\nocite{*}
\bibliography{reference}% Produces the bibliography via BibTeX.

\end{document}